\documentclass[aip,reprint]{revtex4-1}
\usepackage{amsmath}
\usepackage{color}
\usepackage{hyperref}
\usepackage{tabularx}
\usepackage{multirow}
\usepackage{graphicx}
\usepackage[utf8]{inputenc}
\usepackage[T1]{fontenc}
\usepackage{mathptmx}
\usepackage{etoolbox}

\draft 

\begin{document}


\title{Effect of horizontal magnetic field on K\"{u}ppers-Lortz instability} 

\author{Sutapa Mandal}
\affiliation{Department of Mathematics, National Institute of Technology, Durgapur-713209, India}
\author{Snehashish Sarkar}
\affiliation{Department of Mathematics, National Institute of Technology, Durgapur-713209, India}
\author{Pinaki Pal}
\email{pinaki.pal@nitdgp.ac.in}
\affiliation{Department of Mathematics, National Institute of Technology, Durgapur-713209, India}

\date{\today}




\begin{abstract}
We investigate the effect of an external horizontal magnetic field on the K\"{u}ppers-Lortz  instability (KLI) in rotating Rayleigh-B\'{e}nard convection of Boussinesq fluids using weakly nonlinear theory along with linear theory. By KLI, we mean the instability where the two-dimensional roll solutions of the system occurring at the onset of convection becomes unstable against the perturbations by rolls oriented at different angle with the previous one as the rotation rate exceeds a critical value. The governing parameters, namely, the Prandtl number ($\mathrm{Pr}$), Taylor number ($\mathrm{Ta}$) and Chandrasekhar number ($\mathrm{Q}$) are varied in the ranges $0.8 \leq \mathrm{Pr} < \infty$, $0 < \mathrm{Ta} \leq 10^4$ and $0 \leq \mathrm{Q} \leq 10^4$ respectively by considering the vanishingly small magnetic Prandtl number limit. In the $\mathrm{Pr}\rightarrow \infty$ limit, magnetic field is found to inhibit the KLI by enhancing the critical Taylor number ($\mathrm{Ta}_c$) for its onset. On the other hand, for finite Prandtl number fluids,  KLI is favored for lower $\mathrm{Q}$, and it is inhibited for higher $\mathrm{Q}$. Interestingly, in the finite Prandtl number range both KLI and small angle instability are manifested depending on the Prandtl number. No small angle instability is observed for $\mathrm{Pr} \geq 50$ and the rotation induced KLI is inhibited predominantly by the magnetic field. While, for $\mathrm{Pr} < 50$, along with the K\"{u}ppers-Lortz  instability, small angle instability is also observed. However, in this case, KLI is favored for lower $\mathrm{Q}$, while it is inhibited for higher $\mathrm{Q}$. 
\end{abstract}

\pacs{}

\maketitle 

\section{Introduction}\label{sec1:Intro}
The study of thermal convection in the presence of rotation or magnetic field or both has been the topic of profound interest among the researchers for last several decades, primarily due to its relevance in geophysical and astrophysical systems along with its very rich nonlinear dynamics ~\cite{chandra:book,Ecke:ARFM2023,Guzman:JFM928_2021,Schumachar:JFM895_2020,novi:PRE_2019,wang:JFM_2020,hartmann:JFM_2022}. Thus, researchers investigated various aspects of thermal convection, including instabilities, using simplified models like Rayleigh-B\'enard convection (RBC)~\cite{chandra:book,Loshe_RMP:2009,bodenschatz:ARFM_2000,mkv:book}. As a result, notwithstanding the simplicity of the considered geometry, a plethora of interesting instabilities has been discovered which not only enhanced the understanding of the basic physics of convection but also contributed significantly to the development of the subjects like hydrodynamic instabilities and nonlinear dynamics~\cite{Proctor:Book,pal:EPJB_2012,mandal:POF_2022,ghosh:POF_2020,banerjee:PRE_2020,garai:POF_2022}. 

K\"{u}ppers-Lortz  instability (KLI)~\cite{kuppers:1969,Kuppers:PL1970,bodenschatz:ARFM_2000} is one of the fascinating instabilities observed in rotating Rayleigh-B\'enard convection (RRBC).  It occurs close to the onset of convection as the stationary two-dimensional rolls pattern loses stability against the perturbations in the form of rolls of same wave number, making an angle ($\theta_c$) with the previous one for sufficiently fast rotation rate. The relevant control parameters in this case are the Rayleigh number ($\mathrm{Ra}$, vigor of buoyancy), Taylor number ($\mathrm{Ta}$, rotation rate), and Prandtl number ($\mathrm{Pr}$, ratio of thermal and viscous diffusion time scales).  In the first report of KLI~\cite{kuppers:1969},  RRBC of infinite Prandtl number fluids ($\mathrm{Pr}\rightarrow \infty$) with free-slip boundary conditions was considered and using weakly nonlinear theory, it was shown that as the Taylor number ($\mathrm{Ta}$) exceeds a critical value ($\mathrm{Ta}_c = 2285\pm 0.1$), the two-dimensional rolls become unstable with respect to the perturbations in the form of similar rolls oriented at an angle $\phi_c = 58^o \pm 0.5^o$ and  time dependent flow was predicted just at the onset.  In a subsequent theoretical work~\cite{Kuppers:PL1970}   KLI was also reported for finite Prandtl number fluids with realistic no-slip boundary conditions.  It was found that  $\mathrm{Ta}_c$ is greatly reduced with the lowering of {$\mathrm{Pr}$}.  These theoretical works immediately led to the first experimental study on K\"{u}ppers-Lortz  instability~\cite{krishnamurti:1971} and  reported remarkable similarity with the theoretical predictions for  $\mathrm{Pr} > 1$.   

The  time dependent solutions arising out of K\"{u}ppers-Lortz  instability  at the onset of convection eventually give rise to spatio temporal chaos (domain chaos) characterized by the patterns consisting of the domains of rolls constantly replacing each other~\cite{Clever:JFM1979, niemela:PRL_1986}.  This surprising result inspired numerous theoretical~\cite{cross:1994,ponty:POF_1997,podvigina:2010} as well as experimental~\cite{Hu:PRL_1995,hu:PRE_1997,hu:PRE_1998} works in rotating convection for a wide range of fluids mainly focusing on the pattern formation and related time and length scales of the instability near  the onset~\cite{bodenschatz:ARFM_2000}.  However,  unlike the infinite Prandtl number fluids,   KLI of finite $\mathrm{Pr}$ fluids with free-slip boundary conditions is greatly affected by the appearance of horizontal mean flow ~\cite{ponty:POF_1997,Clune:PRE1993}.  Moreover,  in all the works mentioned so far  ignored the role  of centrifugal buoyancy on the KLI which was considered subsequently by Rubio~\cite{rubio:JFM_2010} and reported closer match with the experimental results\cite{Ning:PRE_1993,bajaj:PRE_2002}. On the other hand,  the study of KLI in the presence of an external magnetic field~\cite{jkb_das:1990,podvigina:2010} is extremely limited .   So far K\"{u}ppers-Lortz  instability  in the presence of an external vertical magnetic field using weakly nonlinear  theory has been considered in the literature and reported the inhibitory effect of external magnetic field on it.   

Afterwards, most of the works on rotating convection focused on the  turbulent flow regimes in the absence of magnetic field~\cite{favier:JFM_2019,vishnu:POF_2019,yang:PRF_2020,kumar:POF_2022,Ecke:ARFM2023} except a few recent theoretical works on KLI in nano-fluids~\cite{kanchana:2020} and porous media~\cite{siddheshwar:2020} near the onset.  Also, only a limited attention was given to the study of instabilities and related bifurcation structures near the onset of rotating convection. 
However, in some of the recent works on rotating convection ~\cite{Maity:POF2014,mandal:POF_2022} and magnetoconvection~\cite{Eltayeb:1972,busse:1989a,pal:EPJB_2012,arnab:2014,nandu:2015,ghosh_PRE:2021,mandal:EPL_2021}, extremely rich bifurcation structure has been reported near the onset of convection~\cite{ghosh:2017,banerjee:PRE_2020}.  Interestingly, despite the possibility of a very rich  bifurcation structure near the onset of rotating convection, it is evident from the literature that the effect of a horizontal external magnetic field on KLI has not been explored so far. Note that the primary instability i.e. the onset of convection does not change in presence of a horizontal magnetic field but the nonlinear pattern selection at the onset of convection and secondary instabilities or higher order instabilities are known to be greatly affected by the horizontal magnetic field~\cite{libchaber_et.al:JPL_1982,fauve:1984,ghosh:POF_2020,lekha:PRF2022}. Therefore, the effect of an external horizontal magnetic field on the K\"{u}ppers-Lortz  instability is expected to be very interesting.      

Thus, we revisit the problem of KLI to investigate the effect of an external horizontal magnetic field using weakly nonlinear theory.  Two additional parameters, namely the Chandrasekhar number ($\mathrm{Q}$, strength of the magnetic field) and  magnetic Prandtl number ($\mathrm{Pm}$, ratio of magnetic to viscous diffusion time scales), are required to describe the flow. It is well known that the external horizontal magnetic field enhances the stability regions of the two-dimensional rolls by suppressing the three-dimensionality in the absence of rotation~\cite{chandra:book,pal:EPJB_2012}.  In the presence of rotation, magnetic field also stabilizes the flow by expanding the stability region of the 2D rolls~\cite{ghosh:POF_2020,Eltayeb:2013} for slow rotation rate. However, for high rotation rates, the 2D rolls may become unstable through KLI, as observed in the presence of a vertical magnetic field~\cite{jkb_das:1990}. A relevant question is then how the external  horizontal magnetic field affect the KLI? The present work employs weakly nonlinear theory along with linear theory in a wide region of the parameter space ($0.8 \leq \mathrm{Pr} < \infty$, $0 < \mathrm{Ta} \leq 10^4$ and $0 \leq \mathrm{Q} \leq 10^4$)  considering the vanishingly small magnetic Prandtl number limit to answer this question.   The investigation largely reveals the inhibitory effect of the external magnetic field on KLI, with an exception for low Prandtl number fluids in presence of a weak magnetic field, where KLI is promoted.  Interestingly,  for low Prandtl number fluids, a small angle instability is found to appear along with KLI which dominates for very low Prandtl number fluids. 

The paper is organized as follows: Section~\ref{sec2} describes the mathematical formulation of the problem and it is followed by a discussion on linear theory in the section~\ref{sec3:LT}.  The investigation of KLI using weakly nonlinear theory and obtained results are presented in detail in the section~\ref{sec4:LKI}. Section~\ref{sec5:conclusion} presents general conclusions of the work.

\section{Mathematical formulation}\label{sec2}

\subsection{Physical System}
We wish to investigate the K\"{u}ppers-Lortz instability under the paradigm of the classical plane layer Rayleigh-B\'enard convection (RBC). It consists of an infinitely extended horizontal layer of Bousinesq fluid of thickness $h$, magnetic diffusivity $\lambda$, thermal diffusivity $\kappa$ and kinematic viscosity $\nu$ confined between two perfectly thermally as well as electrically conducting plates in the presence of a horizontal magnetic field ${\bf B_0}=(0,B_0,0)$.  The fluid is subjected to an adverse temperature gradient $\beta=\frac{T_l-T_u}{h}$, where $T_l$ and $T_u$ ($T_l > T_u$) are 
the temperatures at the lower and upper plates respectively. The system is rotated about the vertical axis with an angular velocity ${\Omega}$.  A schematic diagram of the physical system is shown in the FIG.~\ref{RMC}. 
\begin{figure}[h!]
\includegraphics[height=!,width=0.45\textwidth]{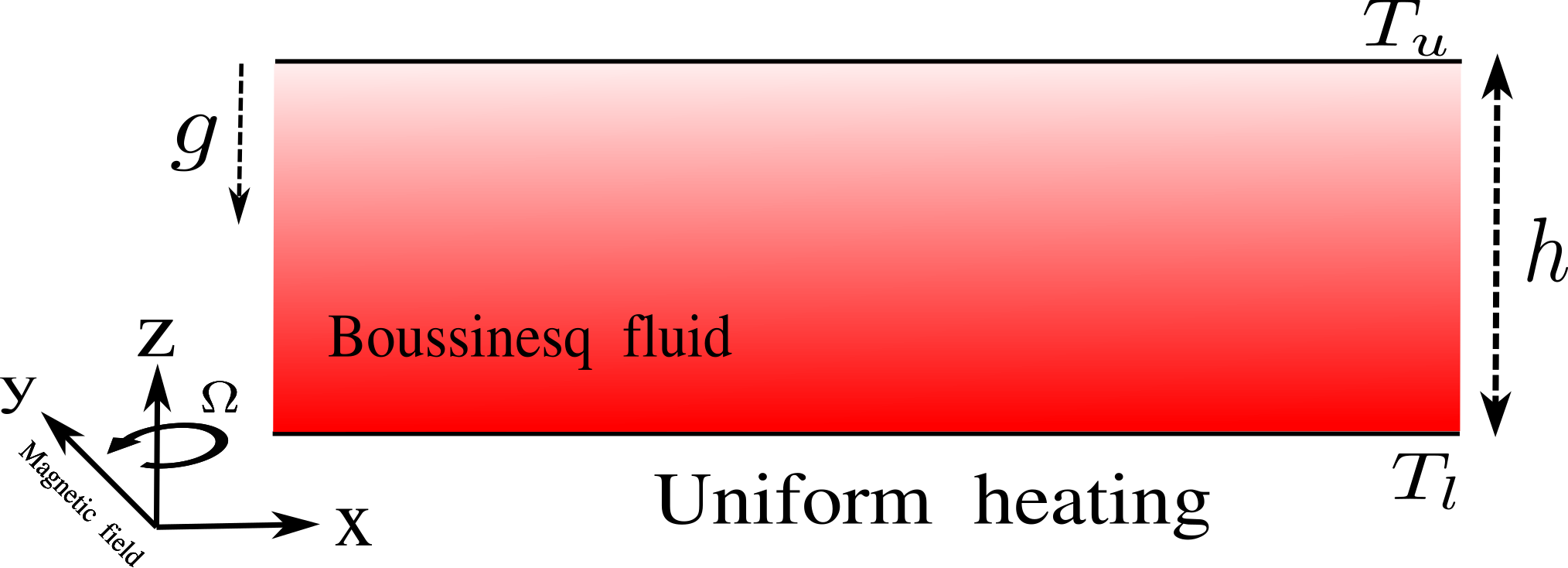}
\caption{Schematic diagram showing the cross sectional view of the rotating magnetoconvection system.}
\label{RMC}
\end{figure}

\subsection{Governing Equations}
The dimensionless governing equations of the hydro-magnetic system under Boussinesq approximation~{\cite{boussinesq:1903,boussinesq_spiegel:1960} with respect to a set of co-ordinate axes fixed at the lower plate and co-rotating with the system are given by
\begin{eqnarray}
\frac{1}{\mathrm{Pr}}\left[\frac{\partial \bf{u}}{\partial t} + (\bf{u}{\cdot}\boldsymbol{\nabla})\bf{u}\right] &=& -\boldsymbol{\nabla}{\Pi} + \nabla^2{\bf{u}} + \mathrm{Ra} \theta {\bf{\hat{e}}}_3 + \sqrt{\mathrm{Ta}}({\bf u} \times {\bf{\hat{e}}}_3)   \nonumber\\ &&+\mathrm{Q}\left[\frac{\partial{\bf b}}{\partial y} + \frac {\mathrm{Pm}}{\mathrm{Pr}}({\bf b}{\cdot}\boldsymbol{\nabla}){\bf b}\right] , \label{eq:momentum} \\   
\mathrm{Pm}[\frac{\partial{\bf b}}{\partial t} + ({\bf u}{\cdot}\boldsymbol{\nabla}){\bf b} &-& ({\bf b}{\cdot}\boldsymbol{\nabla}){\bf u}] = \mathrm{Pr}\left[{\nabla}^2 {\bf b} + \frac{\partial{\bf u}}{\partial y}\right],\label{eq:induction}\\      
\frac{\partial \theta}{\partial t}+(\bf{u}{\cdot}\boldsymbol{\nabla})\theta &=& w+\nabla^2\theta \label{eq:heat},\\
\boldsymbol{\nabla} {\cdot} \bf{u}=0, && \boldsymbol{\nabla} {\cdot} \bf{b}=0, \label{eq:div_free}
\end{eqnarray} 
where ${\bf u}(x,y,z,t)=(u,v,w)$, $\theta(x,y,z,t)$, ${\bf b}(x,y,z,t)=(b_1,b_2,b_3)$,  are velocity, convective temperature and induced magnetic fields, and $\Pi$  is the modified pressure field. The governing equations are made dimensionless using the scales $h$, $\frac{h^2}{\kappa}$, $\frac{\kappa}{h}$, $\beta h$, $\frac{\kappa B_0}{\lambda}$ for length, time, velocity, temperature, and induced magnetic field respectively. The system is now governed by the five dimensionless parameters, namely, the Rayleigh number ($\mathrm{Ra}=\frac{g\alpha \beta h^4}{\nu \kappa}$), the Chandrasekhar number ($\mathrm{Q}=\frac{B_0^2h^2}{\rho_l \nu \lambda}$, $\rho_l$ is the density of the fluid at lower plate), the Taylor number ($\mathrm{Ta}=\frac{4\Omega^2h^4}{\nu^2}$), the thermal Prandtl number ($\mathrm{Pr}=\frac{\nu}{\kappa}$), and the magnetic Prandtl number ($\mathrm{Pm}=\frac{\nu}{\lambda}$). 

\subsection{Boundary Conditions}
The top and bottom plates are assumed to be {\it stress-free} and perfectly thermally conducting, which imply,
\begin{equation}
w=\frac{\partial u}{\partial z}=\frac{\partial v}{\partial z}= \theta = 0 ~\mathrm{at} ~ z=0,1. \label{bc1}
\end{equation}

Further, the plates are also considered electrically conducting, which leads to the conditions 
\begin{equation}
b_3=\frac{\partial b_1}{\partial z}=\frac{\partial b_2}{\partial z}=0 ~ \mathrm{at} ~ z=0,1. \label{bc2}
\end{equation}

In the horizontal directions, all the fields are assumed to be periodic.

\section{Linear Theory}\label{sec3:LT}
Here we employ linear theory~\cite{chandra:book} to determine the critical Rayleigh number ($\mathrm{Ra}_c$) and corresponding wave number ($k_c$) at the onset of convection.  First we take {\it curl} of the equation (\ref{eq:momentum}) twice and equation (\ref{eq:induction}) once to determine  the  following equations of vertical vorticity ($\xi$), velocity ($w$), induced magnetic field ($b_3$) and current density ($j_3$):
\begin{eqnarray}
\left[\frac{1}{\mathrm{Pr}}\frac{\partial}{\partial t} -\nabla^2 \right] \xi &=& \sqrt{\mathrm{Ta}}{\bf D}w +\mathrm{Q}\frac{\partial j_3}{\partial y} - \frac{1}{\mathrm{Pr}} {\bf{\hat{e}}}_3\cdot\left[\boldsymbol\nabla \times ({\bf u}\cdot \boldsymbol\nabla){\bf u}\right]  \nonumber\\&+& \mathrm{Q}{\bf{\hat{e}}}_3\cdot\frac{\mathrm{Pm}}{\mathrm{Pr}}\left[\boldsymbol\nabla \times ({\bf b}\cdot \boldsymbol\nabla){\bf b}\right],\label{eq:vorticity}
\end{eqnarray}

\begin{eqnarray}
\nabla^2 \left[\frac{1}{\mathrm{Pr}}\frac{\partial }{\partial t} - \nabla^2 \right]w = \mathrm{Ra} \nabla^2_H \theta - \sqrt{\mathrm{Ta}}{\bf D} \xi   \nonumber\\+\frac{1}{\mathrm{Pr}} {\bf{\hat{e}}}_3\cdot\left[\boldsymbol\nabla \times\boldsymbol\nabla\times(\bf u \cdot \boldsymbol\nabla)\bf u\right]  \nonumber\\ + \mathrm{Q}{\bf{\hat{e}}}_3\cdot\left[\nabla^2 \frac{\partial{\bf b}}{\partial y} - \frac {\mathrm{Pm}}{\mathrm{Pr}}(\boldsymbol\nabla \times\boldsymbol\nabla\times(\bf b \cdot \boldsymbol\nabla)\bf b\right] , \label{eq:velocity}  
\end{eqnarray}
\begin{equation}
\begin{small}
\left[\frac{\mathrm{Pm}}{\mathrm{Pr}}\frac{\partial }{\partial t} - \nabla^2 \right] b_3 =  \frac{\mathrm{Pm}}{\mathrm{Pr}}{\bf{\hat{e}}}_3\cdot\left[(\bf b \cdot \boldsymbol\nabla) \bf u - (\bf u\cdot \boldsymbol\nabla)\bf b\right]+ \frac{\partial w}{\partial y},
\end{small}
\label{eq:magnetic}
\end{equation}
\begin{equation}
\begin{small}
\left[\frac{\mathrm{Pm}}{\mathrm{Pr}}\frac{\partial }{\partial t} - \nabla^2 \right] j_3 =  \frac{\mathrm{Pm}}{\mathrm{Pr}}{\bf{\hat{e}}}_3\cdot\boldsymbol\nabla \times\left[(\bf b \cdot \boldsymbol\nabla) \bf u - (\bf u\cdot \boldsymbol\nabla)\bf b\right]+ \frac{\partial {\xi} }{\partial y}, \label{eq:current_density}
\end{small}
\end{equation} 
where $\nabla^2_H=\frac{\partial^2}{\partial x^2}+\frac{\partial^2}{\partial y^2}$ is the horizontal Laplacian, ${\bf D}\equiv \frac{d}{dz}$ and ${\bf{\hat{e}}}_3$ is the vertical unit vector. 

For the present investigation we choose the  parameter regime given by $0.8 \leq \mathrm{Pr} < \infty$,  $0 \leq \mathrm{Q} \leq 10^4$ and $0 \leq \mathrm{Ta} \leq 10^4$  for vanishingly small $\mathrm{Pm} $ in such a way that overstable solutions at the onset of convection are avoided.  Next, for performing the linear stability analysis of the motionless conduction state, we drop the nonlinear terms in the equations (\ref{eq:vorticity}) - (\ref{eq:current_density})  and  eliminate $b_3$,  $j_3$,  and $\xi$ from the resulting set of equations.  As a result,  the  following  linearized equation for $w$  
\begin{equation}
\left[(\mathrm{Q}\frac{\partial^2}{\partial y^2}-\nabla^4)(\nabla^6-\nabla^2_H\mathrm{Ra}-\mathrm{Q}\frac{\partial^2}{\partial y^2})-\mathrm{Ta}\nabla^4{\bf D^2}\right] w=0. 
\label{linear_w}
\end{equation}  
is obtained for stationary cellular convection.  We then expand $w$ in terms of normal mode as  
\begin{equation}
w=W(z)e^{i(k_1x+k_2y)+\sigma t}, \label{normal_mode}
\end{equation}
where $k_1$ and $k_2$ are the horizontal wave numbers along the $x$ and $y$ directions respectively and $\sigma$ is the growth rate.  The expression (\ref{normal_mode}) is then substituted in the equation (\ref{linear_w}) together with the boundary condition compatible trial function $W=A\mathrm{sin}\pi z$  to obtain the following expression for  $\mathrm{Ra}$: 
\begin{equation}
\mathrm{Ra} = \frac{(\pi^2+k^2)}{k^2}\left[(\pi^2+k^2)^2+\mathrm{Q}k_2^2\right] + \frac{\mathrm{Ta}\pi^2(\pi^2+k^2)^2}{k^2\left[(\pi^2+k^2)^2+\mathrm{Q}k_2^2\right]}, \label{dispersion_Ra}
\end{equation}
where $k = \sqrt{k_1^2 + k_2^2}.$ The equation (\ref{dispersion_Ra}) is then used to determine the minimum value of $\mathrm{Ra}$ for given $\mathrm{Q}$ by varying $k$ and we obtain the critical Rayleigh number $\mathrm{Ra}_c$ with corresponding critical wave number $k_c$.
\begin{figure}[h!]
\includegraphics[height=!,width=0.5\textwidth]{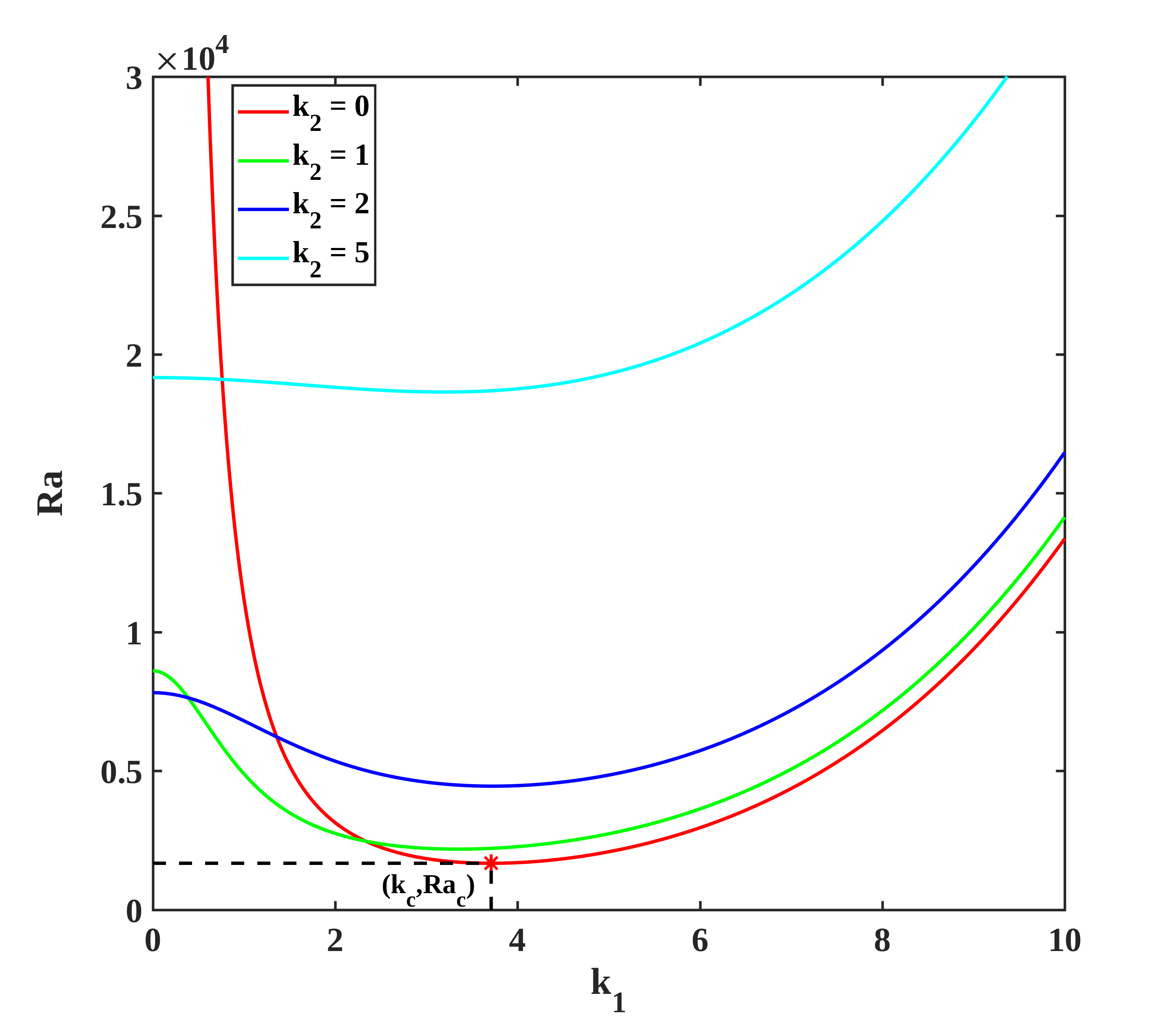}
\caption{Preferred mode of convection at the stationary onset computed from linear theory for $\mathrm{Q}=500$ and $\mathrm{Ta}=1000$.}
\label{RMC}
\end{figure}

The FIG.~\ref{RMC} presenting the variation of $\mathrm{Ra}$ with $k_1$ for different values of $k_2$, shows that the minimum of $\mathrm{Ra}$ occurs for $k_2 = 0$ and $k_1 = 3.71$. Thus, critical wavenumber ($\mathrm{k}_c$) for the onset of convection is $\mathrm{k}_c = k_1 = 3.71$ and the two dimensional (2D) rolls mode considered in the following section is the most unstable mode determined from the linear theory.

Interesting to note here that, the  rotational constraint delays the onset of convection~\cite{chandra:book} in the absence of the magnetic field. On the other hand, the onset of convection also delayed due to the action of the Lorenz force introducing Joule dissipation when the external magnetic field is applied in the vertical direction~\cite{chandra:book,burr2001:POF}. However, in the absence of rotation, for the convection in the form of 2D rolls along $y$-axis, when the horizontal magnetic field is applied along the same direction, it does not apply any electromagnetic force and hence there is no Joule dissipation~\cite{Fauve1981:JPL}. Therefore, in this case, the onset of convection (primary instability) remains unchanged, and it affects only the secondary and higher order instabilities. Now, when both rotation and horizontal external magnetic fields act simultaneously, as considered in the present paper, the primary instability depends only on the rotation. This is because, the 2D rolls convection is the most unstable mode of convection at the onset for which there is no Joule dissipation. However, secondary instabilities like K\"{u}ppers-Lortz Instability and small angle instabilities are greatly affected by the external horizontal magnetic field which are discussed subsequently in detail.

\section{K\"{u}ppers-Lortz Instability}\label{sec4:LKI}
To investigate effect of horizontal magnetic field on the K\"{u}ppers-Lortz instability we use standard weakly nonlinear theory~\cite{nayfeh:2008,glendinning:1994}. For this purpose we write the  equations  (\ref{eq:vorticity}) - (\ref{eq:current_density}) in the following compact form: 
\begin{equation}
L_0X+\Delta\mathrm{R}L_1X = N(X,X)+\frac{\partial }{\partial t}MX, \label{compact_eqn}
\end{equation}
where  $X =  [ w, \xi,  \theta]'$,
\begin{eqnarray}
L_0 &=& \begin{bmatrix} \nabla^4-\mathrm{Q}\frac{\partial^2 }{\partial y^2} & -\sqrt{\mathrm{Ta}}\bf D & \mathrm{R_c}\nabla^2_H \\ \sqrt{\mathrm{Ta}}\bf D\nabla^2 & \nabla^4-\mathrm{Q}\frac{\partial^2 }{\partial y^2} & 0 \\ 1 & 0 & \nabla^2  \end{bmatrix},  \\
	L_1 &=&\begin{bmatrix} 0 & 0 & \nabla_H^2 \\ 0 & 0 & 0\\ 0 & 0 & 0\end{bmatrix}, \\
 N(X,X^{'})&=& \begin{bmatrix} -\frac{1}{\mathrm{Pr}}{\bf{\hat{e}}}_3\cdot [\boldsymbol\nabla \times \boldsymbol\nabla \times ({\bf{u}}\cdot \boldsymbol\nabla) {\bf{u^{'}}}] \\
	\frac{1}{\mathrm{Pr}}{\bf{\hat{e}}}_3 \cdot \nabla ^2[\boldsymbol\nabla \times ({\bf{u}}\cdot \boldsymbol\nabla) {\bf{u^{'}}}] \\
	(\bf u \cdot \boldsymbol\nabla)\theta'
	\end{bmatrix},	\\ 	 
M &=&\begin{bmatrix} 	\frac{1}{\mathrm{Pr}}\nabla^2 & 0 & 0 \\ 0 & \frac{1}{\mathrm{Pr}}\nabla^2 & 0\\ 0 & 0 & 1\end{bmatrix}, \label{M_eq}
\end{eqnarray}
 and $\Delta \mathrm{R}=\mathrm{Ra}-\mathrm{Ra}_c.$ In the next two subsections we investigate KLI both for infinite and finite thermal Prandtl number fluids in the vanishingly small magnetic Prandtl number ($\mathrm{Pm} \ll 1$)  limit. 
\subsection{KLI in Infinite Prandtl number fluids}
To start with, we consider infinite  $\mathrm{Pr}$ fluids with vanishingly small $\mathrm{Pm}$.  Under this assumption,  the matrices $N(X,X)$ and $M$ in the equation (\ref{compact_eqn}) reduce to 
$$N(X,X)= \begin{bmatrix} 0 \\ 0 \\ (\bf u \cdot \boldsymbol\nabla)\theta \end{bmatrix} ~\mathrm{and}~ M =\begin{bmatrix} 0 & 0 & 0 \\ 0 & 0 & 0\\ 0 & 0 & 1\end{bmatrix},$$
while, other terms remain same.  Now to investigate the stability of the 2D-rolls solution very close to the onset of convection, we consider the expansions 
\begin{eqnarray}
\Delta \mathrm{R} &=& \sum_{n=1}^{\infty} \epsilon^n \mathrm{Ra_n} = \epsilon \mathrm{Ra_1} + \epsilon^2 \mathrm{Ra_2} + \dots,\label{ex_deltaR}\\
~\mathrm{and}\nonumber\\
X &=& \sum_{n=1}^{\infty} \epsilon^n \mathrm{X_{n-1}} = \epsilon X_0 + \epsilon^2 X_1 + \epsilon^3 X_2 + \dots, \label{ex_X}
\end{eqnarray}	
where $\epsilon$ is a very small positive number and substitute in the equation (\ref{compact_eqn}).  Then from the resulting equation, equating the terms of different orders in $\epsilon$ to zero we get the following equations:
\begin{eqnarray}
O(\epsilon^0):  L_0X_0 &=& 0,\label{first_X}\\ 
O(\epsilon^1):  L_0X_1&=& N(X_0,X_0) + \mathrm{Ra_1}L_1X_0, \label{second_X}\\
O(\epsilon^2): L_0X_2 &=& N(X_1,X_0) + N(X_0,X_1) \nonumber\\
&-& \mathrm{Ra}_1L_1X_1 - \mathrm{Ra_2}L_1X_0.  \label{third_X}                                                                           
\end{eqnarray}

Solving the zeroth order equation (\ref{first_X}) we obtain the following components of $X_0$ along with horizontal components of velocity $u_0$ and $v_0$. 
\begin{eqnarray}
w_0 &=& \cos {k_cx} \sin {\pi z},\\
u_0 &=& -\frac{\pi}{k_c}\sin {k_cx} \cos {\pi z},\\
v_0 &=& \frac{\sqrt{\mathrm{Ta}}\pi}{k_c(k_c^2+\pi^2)}\mathrm{sin}k_cx\mathrm{cos}\pi z,\\
\xi_0 &=& \frac{\sqrt{\mathrm{Ta}}\pi}{k_c^2+\pi^2}\mathrm{cos}k_cx\mathrm{cos}\pi z,\\
\theta_0 &=& \frac{1}{\pi^2+k_c^2}\mathrm{cos}k_cx\mathrm{sin}\pi z. 
\end{eqnarray}
Now, we take the inner product of the equation (\ref{second_X}) with $X_0$ using the following formula for the inner product $\langle f, g\rangle$ of two functions $f$ and $g$ of $x$, $y$ and $z$ 
\begin{equation}
\langle f,g\rangle = \iiint_D fg \,dx\,dy\,dz,\label{IP}
\end{equation}
where $\mathrm{D} = [0, \frac{2\pi}{k_c}]\times [0, \frac{2\pi}{k_c}]\times [0,1]$ which leads to the result $\mathrm{Ra_1}=0$. Using this, we solve the equation (\ref{second_X}) and obtain the first order solution as
\begin{eqnarray}
u_1&=&0,~v_1=0, ~w_1=0, ~\xi_1 = 0, \\
\theta_1 &=& -\frac{1}{8\pi(\pi^2+k_c^2)}\sin{2\pi z}. \label{X1_sol}
\end{eqnarray}
Proceeding similarly, at the second order we obtain
\begin{equation}
\mathrm{Ra_2} = \frac{\mathrm{Ra_c}}{8(\pi^2+k_c^2)}.
\end{equation}
The flow patterns corresponding to the above mentioned analytically obtained weakly  nonlinear solution is found to be 2D rolls oriented along the $y$-axis. This can be clearly seen from the isotherms at the mid-plane ($z = 0.5$) along with three dimensional velocity field shown in the FIG.~\ref{2D_fllow_pattern}.

\begin{figure*}
\includegraphics[height=!,width=0.7\textwidth]{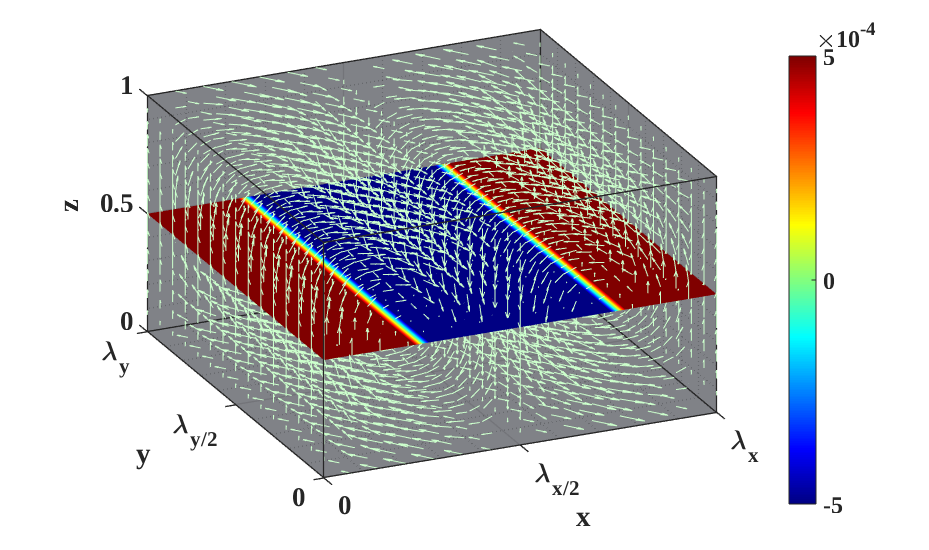}
\caption{Isotherms computed from the expression of $\theta$ (third component of $X$ upto second order term) at the mid-plane $z=0.5$ for $\mathrm{Q}=300$, $\mathrm{Ta}=1000$ and $\epsilon=0.1$. Corresponding velocities are also shown with yellow arrows in the 3D box of dimension $\lambda_x\times\lambda_y \times 1$, where $\lambda_x = \lambda_y = \frac{2\pi}{k_c}$.}
\label{2D_fllow_pattern}
\end{figure*}

Next to investigate the K\"{u}ppers-Lortz instability  we need to check the  stability of the above 2D rolls to another set of rolls making some angle with the previous one.  For that purpose, we consider a perturbation vector $Z= Y e^{pt}$ around the initial 2D rolls solution, where $p$ is the growth rate and use the governing equation (\ref{compact_eqn}) to arrive at the equation
\begin{equation}
L_0Y+\Delta\mathrm{R}L_1Y = N(X,Y)+N(Y,X)+pMY. \label{Y_eqn}
\end{equation}

\begin{figure}[h!]
\includegraphics[height=!,width=0.45\textwidth]{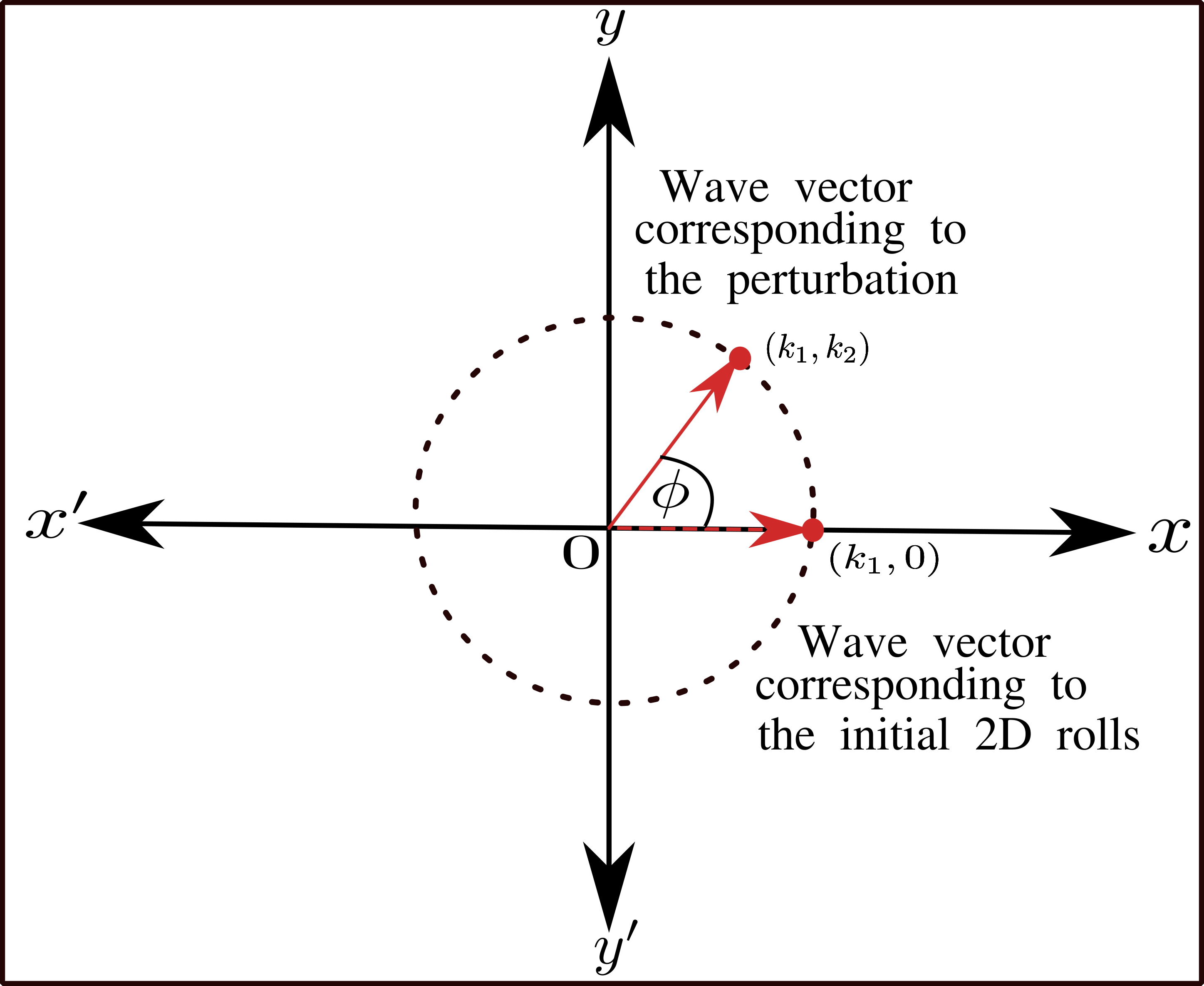}
\caption{Schematic diagram showing the wave vectors corresponding to the initial 2D roll and the perturbation at an angle $\phi$.}
\label{wave_vector}
\end{figure}

Then we consider the series expansions 
\begin{eqnarray}
p &=& \sum_{n=0} ^\infty \epsilon^n p_n = p_0 + \epsilon p_1 + \epsilon^2 p_2 + \dots ,\label{p_expansion}\\
\mathrm{and}\nonumber \\
Y &=& \sum_{n=1} ^\infty \epsilon^n Y_{n-1} = \epsilon Y_0 + \epsilon^2 Y_1 + \dots  \label{Y_expansion}
\end{eqnarray}
and  substitute these along with the perturbation expansions of  $\Delta R$ and $X$ from the equations (\ref{ex_deltaR}) and (\ref{ex_X}) in the equation (\ref{Y_eqn}) which leads to the following equations for different orders of $\epsilon.$ 
\begin{eqnarray}
O(\epsilon^0):  L_0Y_0 &=& p_0MY_0,\label{first_Y}\\ 
O(\epsilon^1):  L_0Y_1&=& N(X_0,Y_0) + N(Y_0,X_0) + p_1MY_0\nonumber\\
                      &+&p_0MY_1-\mathrm{Ra}_1L_1Y_0, \label{second_Y}\\
O(\epsilon^2): L_0Y_2 &=& N(X_1,Y_0) + N(X_0,Y_1) + N(Y_1,X_0)  \nonumber\\
&+& N(Y_0,X_1) - \mathrm{Ra_2}L_1Y_0 - \mathrm{Ra}_1L_1Y_1 + p_2MY_0\nonumber\\
&+& p_0MY_2 + p_1MY_1 .  \label{third_Y}                                                                           
\end{eqnarray}
 We are now interested to determine the marginally stable solution $Y_0$ of the equation (\ref{first_Y}).  Thus, we take $p_0 =0$ and get the equation 
 \begin{equation}
L_0Y_0=0. \label{eq_Y0}
\end{equation}
An exact solution $Y_0 =  [ \tilde{w}_0, \tilde{\xi}_0, \tilde{\theta}_0]'$ of the equation (\ref{eq_Y0}) along with the corresponding horizontal components $\tilde{u}_0$ and $\tilde{v}_0$ are then determined as
\begin{small}
\begin{eqnarray}
\tilde{w_0} &=& {\mathrm{\bf Re}}~e^{i(k_1 x+k_2 y)}\sin{\pi z}, \label{tilde_u0}\\
\tilde{u_0} &=& {\mathrm{\bf Re}}\left[\frac{ik_1\pi}{k_c^2} + \frac{ik_2\sqrt{\mathrm{Ta}}\pi(\pi^2+k_c^2)}{k_c^2[(\pi^2+k_c^2)^2+\mathrm{Q}k_2^2]}\right] e^{i(k_1 x+k_2 y)}\cos {\pi z},\nonumber\\
&~&\\
\tilde{v_0} &=& {\mathrm{\bf Re}}\left[\frac{ik_2\pi}{k_c^2} - \frac{ik_1\sqrt{\mathrm{Ta}}\pi(\pi^2+k_c^2)}{k_c^2[(\pi^2+k_c^2)^2+\mathrm{Q}k_2^2]}\right] e^{i(k_1 x+k_2 y)}\cos {\pi z},\nonumber\\
&~&\\
\tilde{\xi _0} &=& {\mathrm{\bf Re}}~ \left[\frac{\sqrt{\mathrm{Ta}}\pi(\pi^2+k_c^2)}{(\pi^2+k_c^2)^2+\mathrm{Q}k_2^2}\right] e^{i(k_1 x+k_2 y)}\cos{\pi z},\\
\tilde{\theta _0} &=& {\mathrm{\bf Re}}\frac{1}{\pi^2+k_c^2} e^{i(k_1 x+k_2 y)}\sin{\pi z},\label{tilde_theta0}
\end{eqnarray}
\end{small}where $k_1 = k_c\cos{\phi}$ and $k_2 = k_c\sin{\phi}$.  For such choice of $k_1$ and $k_2$,  the above solution represents straight rolls flow patterns inclined at an angle $\phi$ with the previous one. The schematic diagram (FIG.~\ref{wave_vector}) shows the directions of the wave vectors corresponding to the initial 2D rolls and the perturbation in the form of rolls inclined at an angle $\phi$ with the previous one. The weakly nonlinear stability analysis of the initial 2D rolls flow pattern is then performed by finding first and second order nonlinear corrections of the solution $Y_0$ and looking at the sign of the growth rate $p$.  Applying the result $p_0=\mathrm{Ra}_1 =0$ and using the solvability condition $\langle Y_1,Y_0\rangle = 0$ we get $p_1=0$ from the inner product of the first order equation (\ref{second_Y}) with $Y_0$. 
\begin{figure*}
\includegraphics[height=!,width=0.7\textwidth]{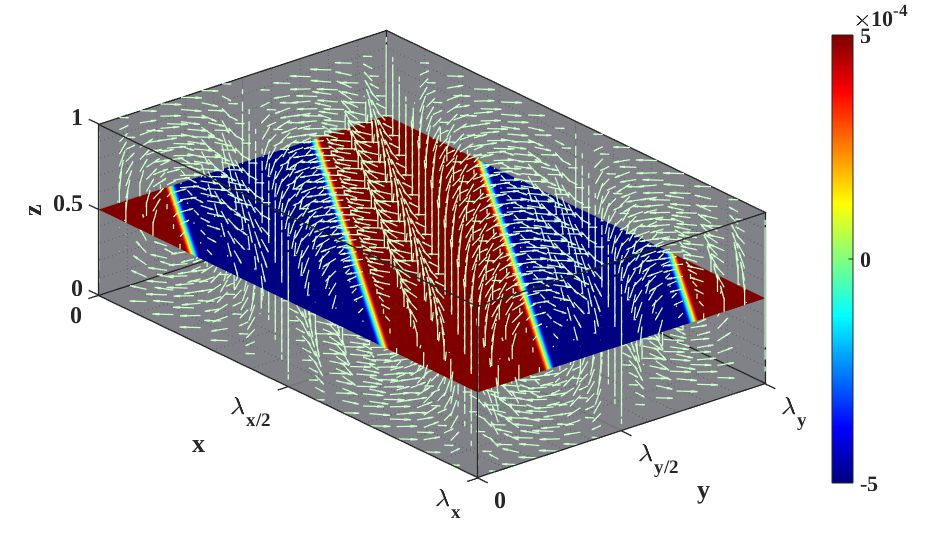}
\caption{Isotherms representing the flow pattern at the onset of KLI. It is computed using the third component of $Y$ at the midplane ($z=0.5$) considering upto second order terms for $\mathrm{Q}=300$, $\mathrm{Ta}=1000$, $\mathrm{Pr}=10$ and $\epsilon=0.1$. Corresponding velocities are also shown with yellow arrows in the 3D box of dimension $\lambda_x\times\lambda_y \times 1$.}
\label{oblique_flowpattern}
\end{figure*}

As a result, first order corrections to $Y_0$ is then determined as 
\begin{widetext}
\begin{eqnarray}
\tilde{w_1} &=& \mathrm{{\bf Re}}~\left[A_+ e^{i(k_1+k_c)x} + A_- e^{i(k_1-k_c)x}\right] e^{ik_2 y}\mathrm{sin}2\pi z,\label{Y1_u0}\\
\tilde{\xi _1} &=& \mathrm{{\bf Re}}~\left[B_+ e^{i(k_1+k_c)x} + B_- e^{i(k_1-k_c)x}\right] e^{ik_2 y}\mathrm{cos}2\pi z,\\
\tilde{\theta _1} &=& \mathrm{{\bf Re}}~\left[C_+ e^{i(k_1+k_c)x} + C_- e^{i(k_1-k_c)x}\right] e^{ik_2 y}\mathrm{sin}2\pi z,\\
\tilde{u_1} &=& \mathrm{{\bf Re}}~\left[\frac{2\pi i(k_1+k_c)A_+ + ik_2B_+}{(k_1+k_c)^2+k_2^2} e^{i(k_1+k_c)x} + \frac{2\pi i(k_1-k_c)A_- + ik_2B_-}{(k_1-k_c)^2+k_2^2} e^{i(k_1-k_c)x}\right] e^{ik_2 y}\mathrm{cos}2\pi z,\\
\tilde{v_1} &=& \mathrm{{\bf Re}}~\left[\frac{2\pi i(k_1+k_c)A_+ + ik_2B_+}{(k_1+k_c)^2+k_2^2} e^{i(k_1+k_c)x} + \frac{2\pi i(k_1-k_c)A_- + ik_2B_-}{(k_1-k_c)^2+k_2^2} e^{i(k_1-k_c)x}\right] e^{ik_2 y}\mathrm{cos}2\pi z,\label{Y1_theta0}
\end{eqnarray}
where  ${\bf Re}$ stands for the real part and the expressions for $A_\pm$, $B_\pm$, and $C_\pm$ are given in the APPENDIX. 
\end{widetext}

\begin{figure*}
\includegraphics[height=!,width=1\textwidth]{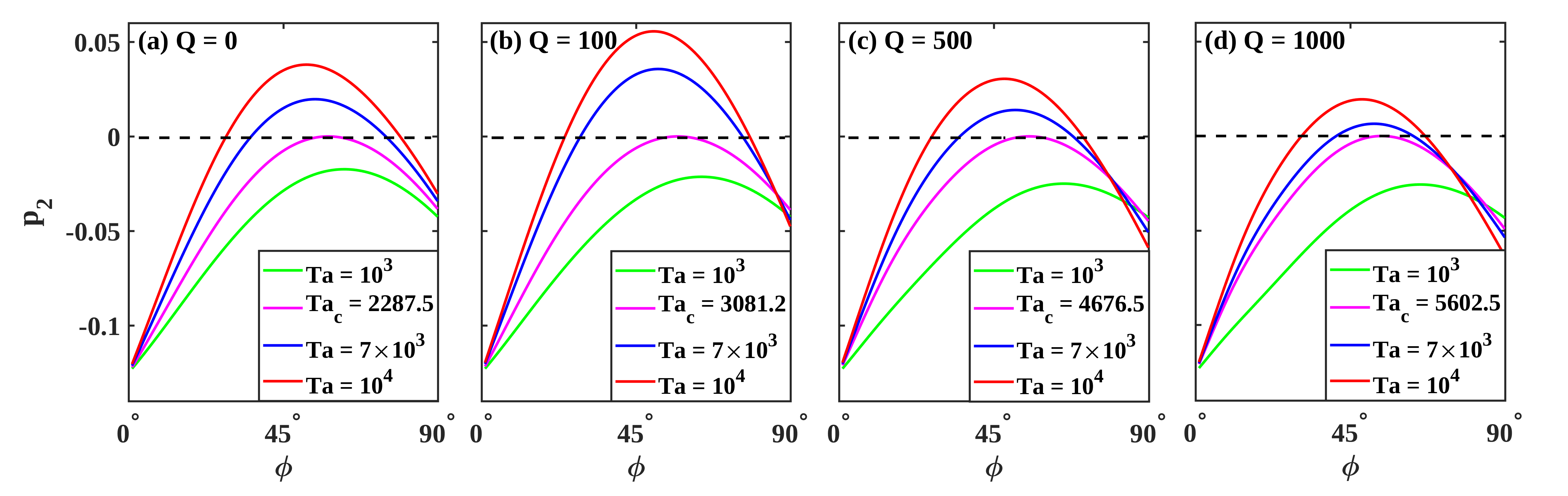}
\caption{Effective growth rate $p_2$ as a function of $\phi$ for four different $\mathrm{Ta}$ including the critical one ($\mathrm{Ta}_c$) in each cases of (a) $\mathrm{Q} = 0$, (b) $\mathrm{Q} = 100$, (c) $\mathrm{Q} = 500$ and (d) $\mathrm{Q} = 1000$ showing the impact of magnetic field on critical Taylor number  and angle for the onset of KL-instability in the infinite Prandtl number limit. The critical Taylor number clearly increases with the strength of the magnetic field indicating the inhibitory effect of it on KLI.}
\label{p2_Q0_1000}
\end{figure*}

Next we apply the solvability conditions $\langle Y_1,Y_0 \rangle = \langle Y_2,Y_0 \rangle = 0$ together with $p_0=\mathrm{Ra}_1 = p_1=0$  in the inner product of the second order equation (\ref{second_Y}) with $Y_0$ and obtain the following expression of the effective growth rate $p_2$
\begin{widetext}
\begin{equation}
p_2 = \frac{\mathrm{sin}\phi}{8}(\frac{B_+}{1+\mathrm{cos}\phi}-\frac{B_-}{1-\mathrm{cos}\phi})-\frac{\pi}{4}[(\mathrm{cos}\phi-1)C_+-(\mathrm{cos}\phi+1)C_-](\pi^2+k_c^2)+\frac{\pi \sqrt{\mathrm{Ta}}\mathrm{sin}\phi}{4}(C_+-C_-).\label{p2_ex}
\end{equation}
\end{widetext}

The above expression of $p_2$ is a function of the parameters $\mathrm{Ra}_c$, $\mathrm{Q}$, $\mathrm{Ta}$, $k_c$ and $\phi$. The weakly  nonlinear stability of the approximate 2D rolls (along $y$-axis) solution is then determined by the sign of $p_2$. The solution $X$ will become unstable when $p_2 > 0$ and the straight rolls solution making an angle $\phi$ with the previous one will start to grow which is marked as the onset of K\"{u}ppers-Lortz instability. The FIG.~\ref{oblique_flowpattern} shows the isotherms at the mid-plane ($z = 0.5$) with corresponding three dimensional velocity field which clearly represents an oblique rolls flow patterns associated with this instability. 

\subsubsection{Numerical Results for infinite $\mathrm{Pr}$}
The onset of KLI in the parameter regime of our interest is now explored in detail in the infinite Prandtl number limit by numerically evaluating $p_2$ from the equation (\ref{p2_ex}). In the process, first we set a fixed value of $\mathrm{Q}\in [0, 10^4]$.  Then for each value of $\phi_c$ in the range $[0, \frac{\pi}{2}]$ we vary the parameter $\mathrm{Ta}$ from $0$ to $10^5$ and note the point where $p_2$ changes sign from negative to positive. 

FIG.~\ref{p2_Q0_1000} shows the variation of $p_2$ with $\phi$ for different pairs of values of the parameters $\mathrm{Ta}$ and $Q$. It is seen from the FIG.~\ref{p2_Q0_1000}(a) that for $\mathrm{Q}=0$, as the Taylor number is increased, the graphs of $p_2$ touches the zero line at $\mathrm{Ta}=2287.5$, marking the critical values $\mathrm{Ta_c}$ and $\phi_c$ for the onset of KLI. For further increase in $\mathrm{Ta}$, the graph crosses the zero line, indicating the range of $\phi$ for which straight 2D rolls oriented along the $y$-axis is unstable. It is observed that with the increment of $\mathrm{Ta}$, this range of $\phi$ increases. The effect of magnetic field on these graphs of $p_2$ can be seen from the FIG.~\ref{p2_Q0_1000}(b)-(d). It is clearly seen that the critical value for the onset of KLI, $\mathrm{Ta}_c$ increases with $\mathrm{Q}$.  

\begin{figure}
\includegraphics[height=!,width=0.5\textwidth]{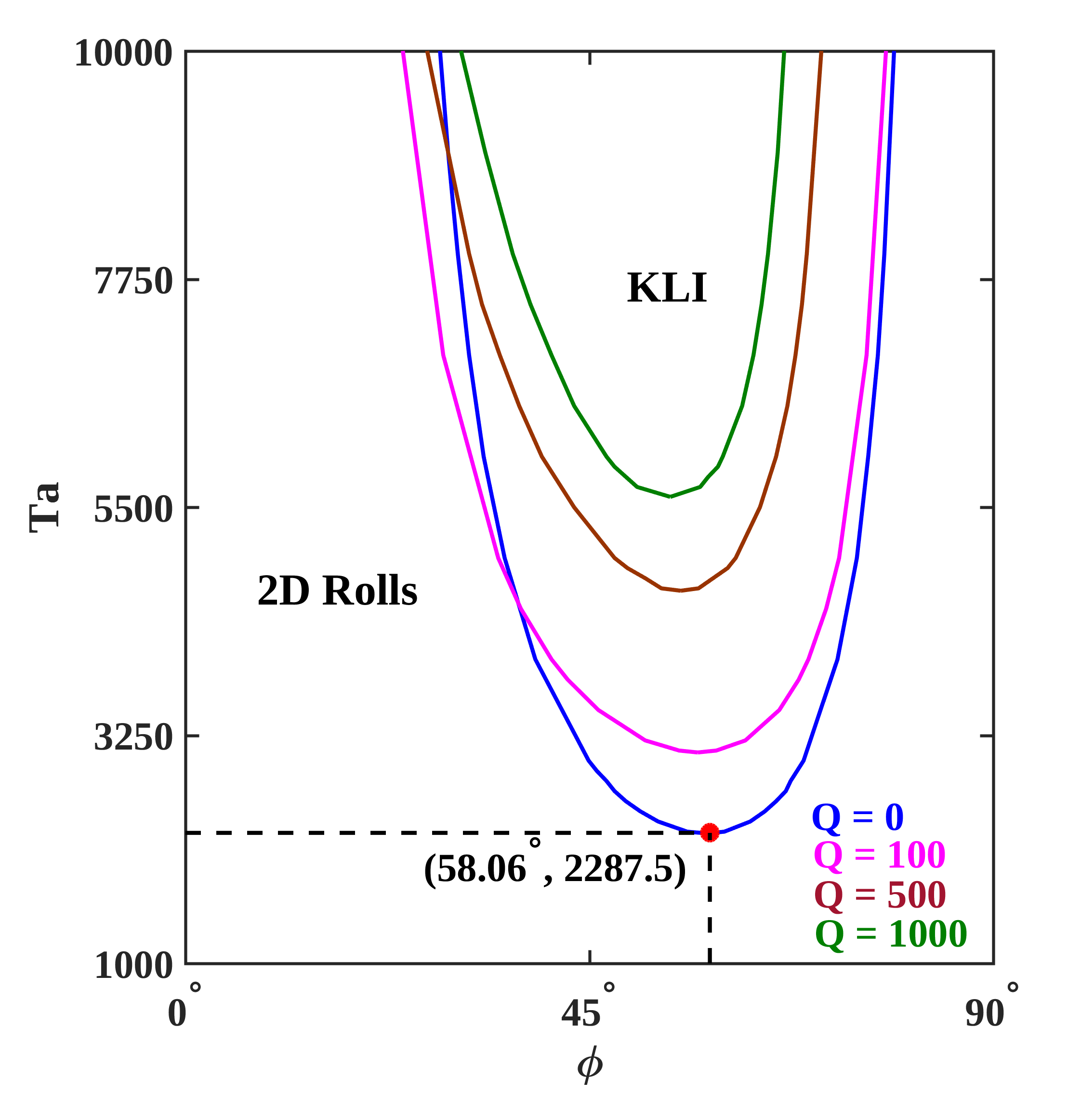}
\caption{Effect of magnetic field on critical Taylor number  and angle for the onset of KL-instability in the infinite Prandtl number limit. The minima of each of the curves shows ($\phi_c$, $\mathrm{Ta}_c$) corresponding to a $\mathrm{Q}$. The minima clearly moves up and slightly shifts towards left as the strength of the magnetic field is increased.}
\label{Ta-theta}
\end{figure}

For more detailed understanding of the parameter space we determine a curve to mark the boundary of the KLI in the $\phi - \mathrm{Ta}$ plane for different $\mathrm{Q}$ from the equation (\ref{p2_ex}). FIG.~\ref{Ta-theta} shows the curves delimiting the stability regions of KLI and 2D rolls along $y$ axis. Each of the curves shows a clear minima which corresponds to the critical Taylor number ($\mathrm{Ta}_c$) and the associated critical angle ($\phi_c$). For $\mathrm{Q} = 0$, the critical values of $\mathrm{Ta}_c$ and $\phi_c$ are found to be $2287.5$ and $58.06^\circ$ which is very close to the one found in the first report of the KLI~\cite{kuppers:1969}. In the presence of magnetic field, the minima is clearly found to move vertically upward and at the same time move towards left in the horizontal direction. Moreover, the area of the KLI region consistently reduced with the increase of the strength of the magnetic field. Therefore, it is evident that in the presence of external magnetic field, the KLI is clearly inhibited for infinite Prandtl number fluids and the stability region of the 2D rolls along the direction of the external magnetic field is greatly enhanced. 

To understand the effect of external magnetic field on the onset of KLI, we vary $\mathrm{Q}$ in the range $[0 - 10^4]$ and determine critical Taylor number ($\mathrm{Ta}_c$) and the critical angle ($\phi_c$) at the onset of KLI using the equation (\ref{p2_ex}). The results are shown in the FIG.~\ref{Tac_thetac_Pinf}. 
\begin{figure}
\includegraphics[height=!,width=0.5\textwidth]{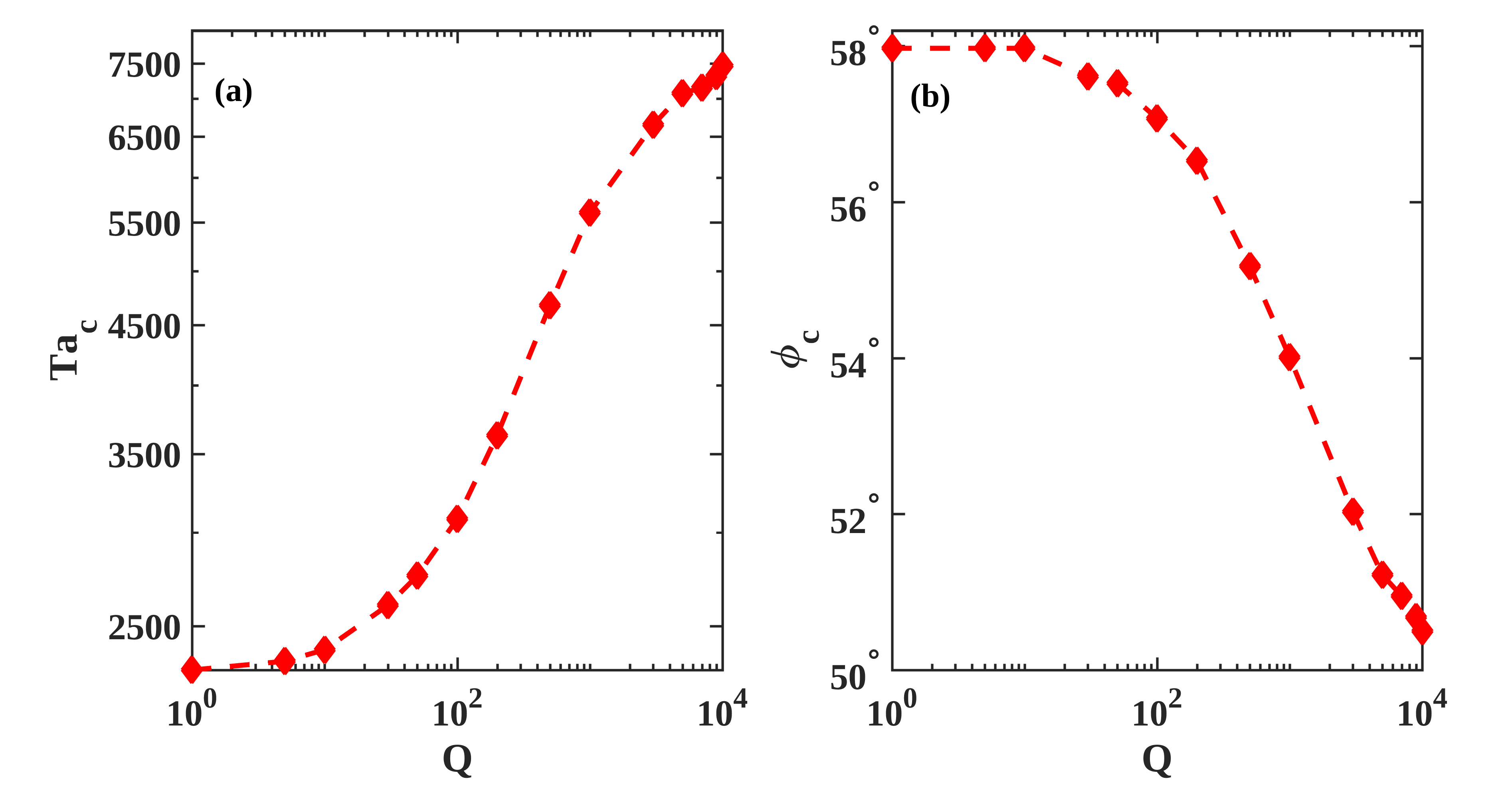}
\caption{Critical (a) Taylor number  and (b) angle for the onset of KL-instability in the infinite Prandtl number limit as a function of $\mathrm{Q}$.}
\label{Tac_thetac_Pinf}
\end{figure}
As expected in the previous paragraph, in infinite Prandtl number fluids, the external magnetic field pushes the boundary for the onset of KLI (FIG.~\ref{Tac_thetac_Pinf}(a)) towards higher Taylor number. Alternatively, horizontal magnetic field inhibits KLI and enhance the stability zone of the initial 2D rolls. Further, from the FIG.~\ref{Tac_thetac_Pinf}(b), it is clear that the corresponding critical angle ($\phi_c$) of the rolls decrease with external horizontal magnetic field. We now move ahead to investigate the effect of external horizontal magnetic field on the KLI in finite Prandtl number fluids. The details are discussed in the next subsection. 

\subsection{Finite Prandtl number}
In this subsection we consider finite $\mathrm{Pr}$ fluids with vanishingly small $\mathrm{Pm}.$ Under these assumptions,  the system is governed by the equations (\ref{compact_eqn})-(\ref{M_eq}).   In this case also the system allow 2D rolls solution given by $\epsilon X_0 + \epsilon^2 X_1$, where $X_0$ is same as the one derived in the previous subsection and $X_1$ together with the horizontal components of the velocity are as follows:
\begin{small}
\begin{eqnarray}
u_1&=&0,~v_1=\frac{1}{\mathrm{Pr}}\frac{\sqrt{\mathrm{Ta}}\pi ^2}{8k^3(\pi ^2+k_c^2)}\sin 2k_cx, ~w_1=0,  \\
\xi_1 &=& \frac{1}{\mathrm{Pr}}\frac{\sqrt{\mathrm{Ta}}\pi ^2}{4k_c^2(\pi ^2+k_c^2)}\cos 2k_cx,~\theta_1 = -\frac{1}{8\pi(\pi^2+k_c^2)}\sin{2\pi z}. \nonumber\\\label{X1_sol_Pr_non}
\end{eqnarray}
\end{small}
As before, the stability of this 2D rolls solution is determined against the convection rolls oriented at an angle $\phi$ with it.  For investigating KLI, the stability analysis of the initial 2D rolls is done by looking at the growth rate $p$ of a perturbation $Z = Y e^{pt}$ which represents a convection roll oriented at an angle $\phi$ with the previous one.  The equation (\ref{Y_eqn}) is then derived using the equation (\ref{compact_eqn}).  Note that in this case the  vector $N$  and the matrix $M$ representing nonlinear  and time dependent terms of the governing equations are different than the infinite Prandtl number case.  Next considering the expansions (\ref{p_expansion}}) and (\ref{Y_expansion}) of  $p$ and $Y$, we derive the equations (\ref{first_Y}) - (\ref{third_Y}) at different orders of $\epsilon.$ Using  the zeroth order equation,  we derive the solution  $Y_0$ given by the equations  (\ref{tilde_u0}) - (\ref{tilde_theta0}) which represents convection rolls oriented at an angle $\phi$ with the previous one.  Next we use the expressions of $X_0$,  $X_1$ and $Y_0$ in the equation (\ref{second_Y}) and determine the first nonlinear correction $Y_1$ given by 

\begin{widetext}
\begin{eqnarray}
\tilde{w_1} &=& \mathrm{{\bf Re}}~\left[e_+ e^{i(k_1+k_c)x} + e_- e^{i(k_1-k_c)x}\right]e^{ik_2y} \sin 2\pi z,\label{finitePr_w1}\\
\tilde{\xi _1} &=& \mathrm{{\bf Re}}~\left[(g_+ + h_+)e^{i(k_1+k_c)x} + (g_- + h_-)e^{i(k_1-k_c)x}\right]e^{ik_2y} \cos 2\pi z,\\
\tilde{\theta_1} &=& \mathrm{{\bf Re}}~\left[f_+ e^{i(k_1+k_c)x} + f_- e^{i(k_1-k_c)x}\right]e^{ik_2y} \sin 2\pi z,\\
\tilde{u_1} &=& \mathrm{{\bf Re}}~\left[\frac{ik_2 h_+ + 2\pi i(k_1+k_c)e_+}{2k_c^2(1 + \mathrm{cos\phi})}e^{i(k_1+k_c)x} + \frac{ik_2 h_- + 2\pi i(k_1-k_c)e_-}{2k_c^2(1 - \mathrm{cos\phi})}e^{i(k_1-k_c)x}\right]e^{ik_2y}\cos \pi z \nonumber\\
 &+& \left[\frac{ik_2g_+}{2k_c^2(1 + \mathrm{cos}\phi)}e^{i(k_1+k_c)x} + \frac{ik_2g_-}{2k_c^2(1 - \mathrm{cos}\phi)}e^{i(k_1-k_c)x}\right]e^{ik_2y},\\
 \tilde{v_1} &=& \mathrm{{\bf Re}}~\left[\frac{i(k_1+k_c)h_+ - 2\pi ik_2e_+}{2k_c^2(1 + \mathrm{cos}\phi)}e^{i(k_1+k_c)x} + \frac{i(k_1-k_c)h_- - 2\pi ik_2e_-}{2k_c^2(1 - \mathrm{cos}\phi)}e^{i(k_1-k_c)x}\right]e^{ik_2y}\cos \pi z \nonumber\\
 &+& \left[\frac{i(k_1+k_c)g_+}{2k_c^2(1 + \mathrm{cos}\phi)}e^{i(k_1+k_c)x} + \frac{i(k_1-k_c)g_-}{2k_c^2(1 - \mathrm{cos}\phi})e^{i(k_1-k_c)x}\right]e^{ik_2y},
\end{eqnarray}
where the expression for the coefficients $q_\pm$, $e_\pm$, $f_\pm$, $g_\pm$ and $h_\pm$ are given in the APPENDIX. 
\end{widetext} 
Then using the solvability conditions $\langle Y_2,  Y_0\rangle = 0$ and $\langle Y_1, Y_0\rangle = 0$ in the equation (\ref{third_Y}) we obtain the following expression for the effective growth rate $p_2$.
\begin{widetext}
\begin{eqnarray}
p_2 &=& -\frac{k_c}{4}\left[2(q_+ -q_-)-(r_+ - r_- + s_+ -s_-)\right] -\frac{\pi}{4}(e_+ + e_-) -\frac{\pi(1+\mathrm{cos}\phi)}{4}f_+ - \frac{\pi(1-\mathrm{cos}\phi)}{4}f_- + \frac{\pi\sqrt{\mathrm{Ta}}~\mathrm{sin}\phi}{4}(f_+ - f_-) \nonumber \\ &+& \frac{\pi (\pi^2 +k_c^2)}{2}(f_+ +f_-),\label{p2_finitePr}
\end{eqnarray}
where the expression of the coefficients $r_\pm$ and $s_\pm$ are given in the APPENDIX. 
\end{widetext}
\subsubsection{Numerical Results for finite $\mathrm{Pr}$}
We now use the equation (\ref{p2_finitePr}) to determine the critical Taylor number for the onset of KLI. First we verify the infinite Prandtl number limit in the absence of external magnetic field ($\mathrm{Q} = 0$). For that purpose, we set $\mathrm{Q} = 0$ and determine $\mathrm{Ta}_c$ and corresponding $\phi_c$ for Prandtl numbers taken in the range $0.8\leq \mathrm{Pr} \leq 10^4.$
\begin{figure}[h!]
\includegraphics[height=!,width=0.5\textwidth]{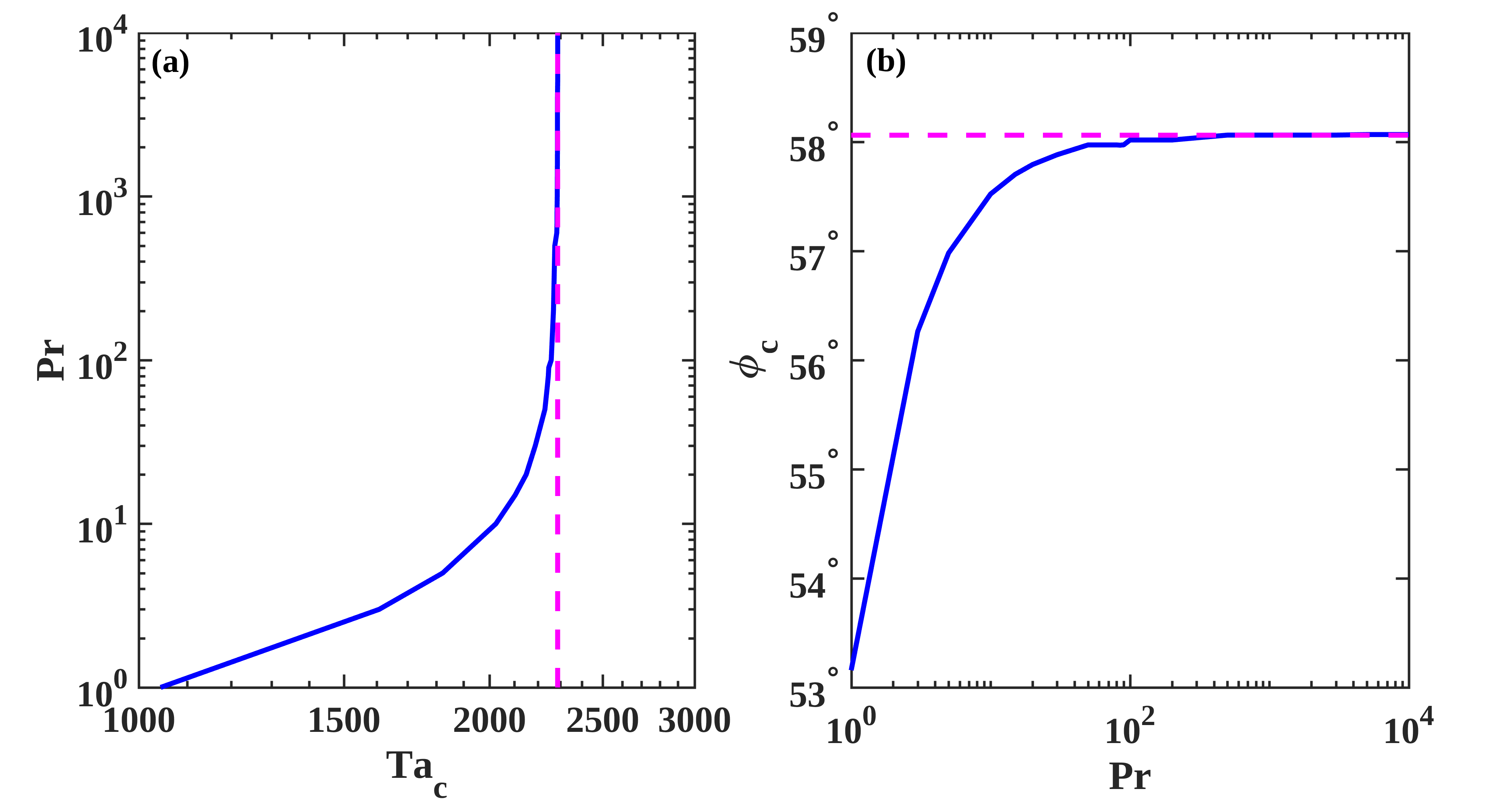}
\caption{K\"uppers-Lortz instability boundary in the $(\mathrm{Ta_c},\mathrm{Pr})$ plane in absence of magnetic field (a) and critical angle for the onset of KL-instability (b). Pink dotted line represent the critical $\mathrm{Ta_c}$ and $\phi_c$ in the infinite $\mathrm{Pr}$ limit in figure (a) and (b) respectively.}
\label{infite_Pr_limit_Q0}
\end{figure}
\begin{table}[htb]
\caption{Critical values of different parameters at the onset of KLI in the absence of magnetic field ($\mathrm{Q} = 0$) are shown as a function of $\mathrm{Pr}$.}\label{table:comparison_of_Rac}
\begin{tabularx}{0.48\textwidth} { 
   >{\centering\arraybackslash}X 
   >{\centering\arraybackslash}X 
   >{\centering\arraybackslash}X
   >{\centering\arraybackslash}X
   >{\centering\arraybackslash}X
   }
 \hline
 \hline
 $\mathrm{Pr}$ & $\mathrm{\tau_c}(\sqrt{\mathrm{Ta_c}})$ & $\phi_c$ & $k_c$ & $\mathrm{Ra_c}$ \\
\hline
     $\infty$  & $47.82$  &  $58.06$  & $4.33$ & $2454$ \\
     $500$  & $47.69$  &  $58.06$ & $4.32$ & $2447$  \\
     $100$  & $47.52$  &  $58.01$ & $4.32$ & $2439$ \\
     $50$ & $47.23$  &  $57.91$ & $4.31$ & $2424$ \\
     $10$  & $45.00$  &  $57.52$ & $4.23$ & $2312$  \\
     $5$  & $42.69$  &  $56.98$ & $4.15$ & $2199$ \\
     $2$  & $37.52$  &  $55.44$ & $3.95$ & $1949$ \\
     $1$  & $32.31$  &  $53.15$ & $3.74$ & $1707$ \\
     $0.8$  & $30.49$  &  $52.06$ & $3.66$ & $1625$ \\
\hline
\end{tabularx}
\end{table}
The graphs prepared from the resulting data are shown in the FIG.~\ref{infite_Pr_limit_Q0}. The FIG.~\ref{infite_Pr_limit_Q0}(a) clearly shows that as $\mathrm{Pr}$ becomes very large, $\mathrm{Ta}_c$ asymptotically approaches towards $\mathrm{Ta}_c = 2287.5$, the critical value for the onset of KLI in the $\mathrm{Pr}\rightarrow \infty$ limit. Also the critical angle ($\phi_c$) at the onset of KLI asymptotically approaches towards $\phi_c = 58.06^\circ$ (FIG.~\ref{infite_Pr_limit_Q0}(b)), the critical value for the onset of KLI in the infinite Prandtl number limit.  Therefore, it is apparent from the figure that the infinite Prandtl number limit is smoothly achieved in the absence of magnetic field. Interesting to note here that $\mathrm{Ta}_c$ decreases with $\mathrm{Pr}$ in the absence of the magnetic field  which is consistent with the theoretical result obtained for free-slip and no-slip boundary conditions~\cite{Clune:PRE1993,ponty:POF_1997,Ponty:PRE1997}. 

For comparing our results with the existing ones in the absence of external magnetic field, we present the critical values of different parameters at the onset of KLI in the table~\ref{table:comparison_of_Rac} as a function of $\mathrm{Pr}$. The tabular data we observe that the critical Taylor number ($\mathrm{Ta}_c$) along with $\phi_c$, $k_c$ and $\mathrm{Ra}_c$ consistently decreases with $\mathrm{Pr}$ which shows close similarity with the results of the theoretical works~\cite{Clune:PRE1993,ponty:POF_1997}.
 
\begin{figure}[h!]
\includegraphics[height=!,width=0.5\textwidth]{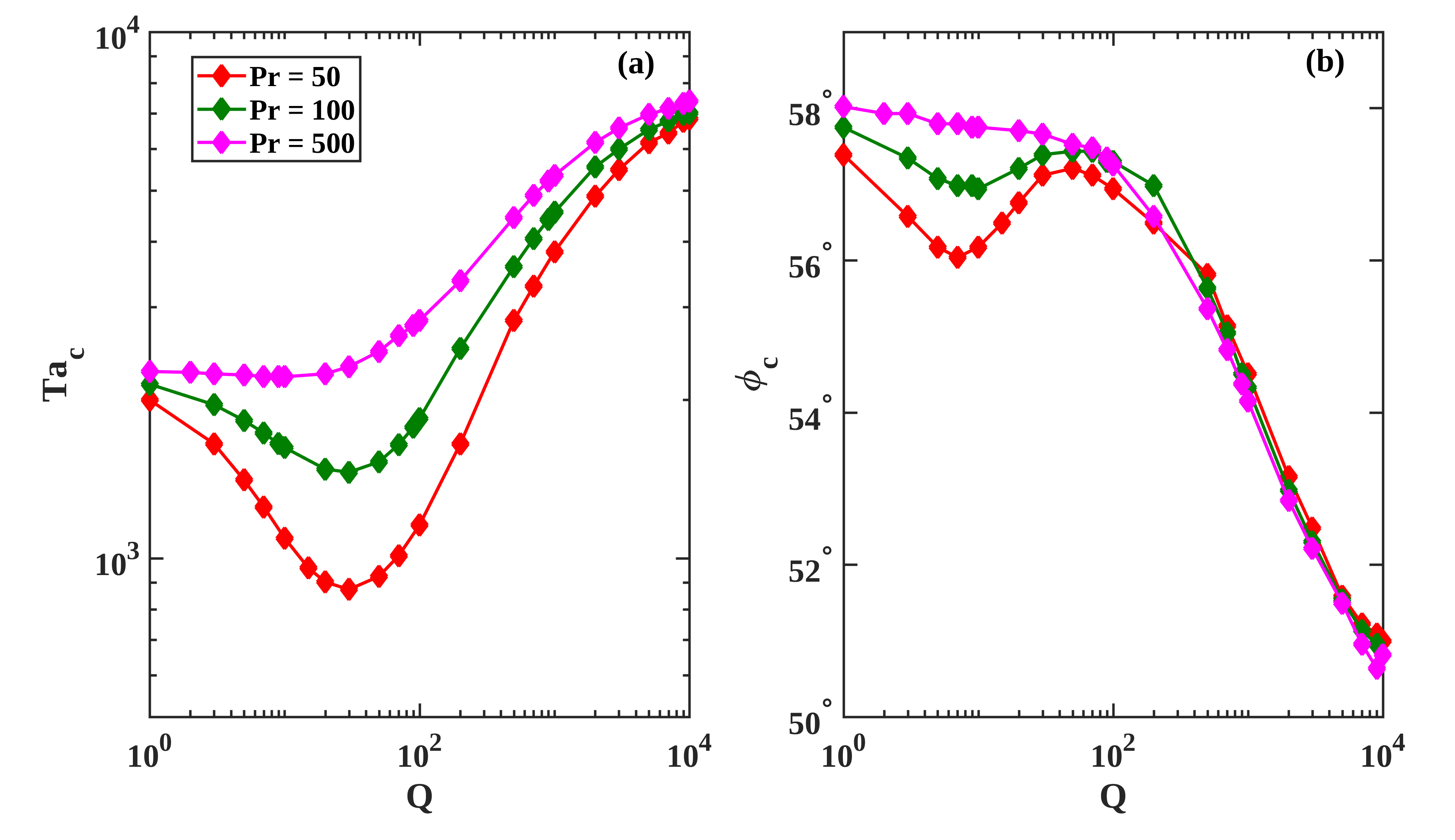}
\caption{(a) Critical Taylor number ($\mathrm{Ta}_c$) and (b) critical angle ($\phi_c$) of the onset of KLI as a function of $\mathrm{Q}$ for large Prandtl numbers.}
\label{largePr_Q}
\end{figure}
\begin{figure}[h!]
\includegraphics[height=!,width=0.5\textwidth]{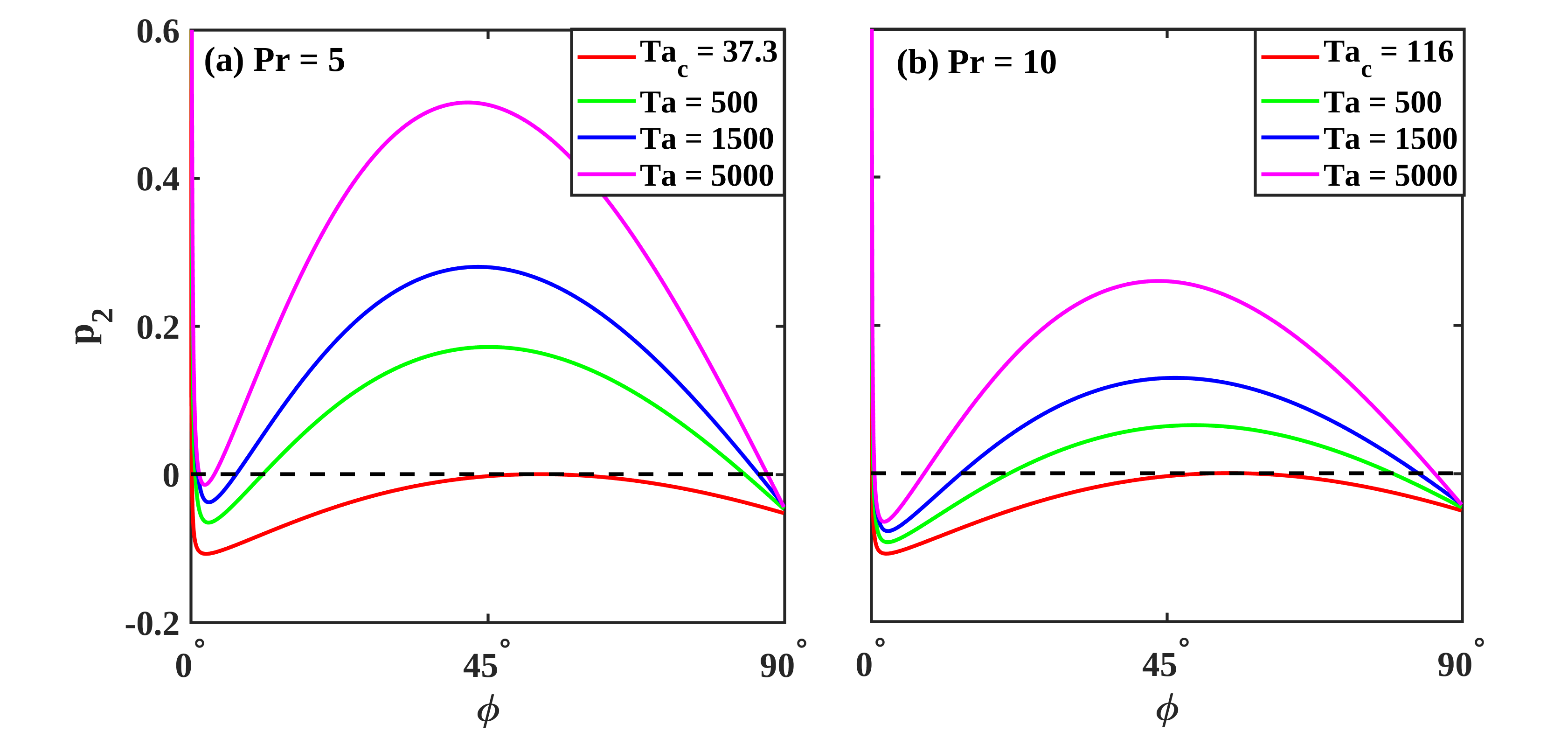}
\caption{Effect of $\mathrm{Ta}$ on the variation of the growth rate $p_2$ with $\phi$ for  $\mathrm{Q} = 50$ with (a) $\mathrm{Pr} = 5$ and (b) $\mathrm{Pr} = 10$. }
\label{smallPr_Q50}
\end{figure}

Next we explore the effect of external horizontal magnetic field on the critical Taylor number and angle at the onset of KLI for large Prandtl number fluids. In the FIG.~\ref{largePr_Q} the variation of $\mathrm{Ta}_c$ and $\phi_c$ with $\mathrm{Q}$ have been shown for three relatively large Prandtl number fluids ($\mathrm{Pr} = 50, 100, 500$). In this case, the graphs of $\mathrm{Ta}_c$ and $\phi_c$ show qualitative similarity with the graphs shown in the FIG.~\ref{Tac_thetac_Pinf} for infinite Prandtl number fluids, except the appearance of local minima near the lower values of $\mathrm{Q}$ for $\mathrm{Pr} = 50$ and $100$.  Match with the infinite Prandtl number case is even better for $\mathrm{Pr} = 500.$ Therefore, for large Prandtl number fluids, horizontal magnetic field largely inhibit the KLI as it has been observed for infinite Prandtl number fluids. 
\begin{figure}[h!]
\includegraphics[height=!,width=0.5\textwidth]{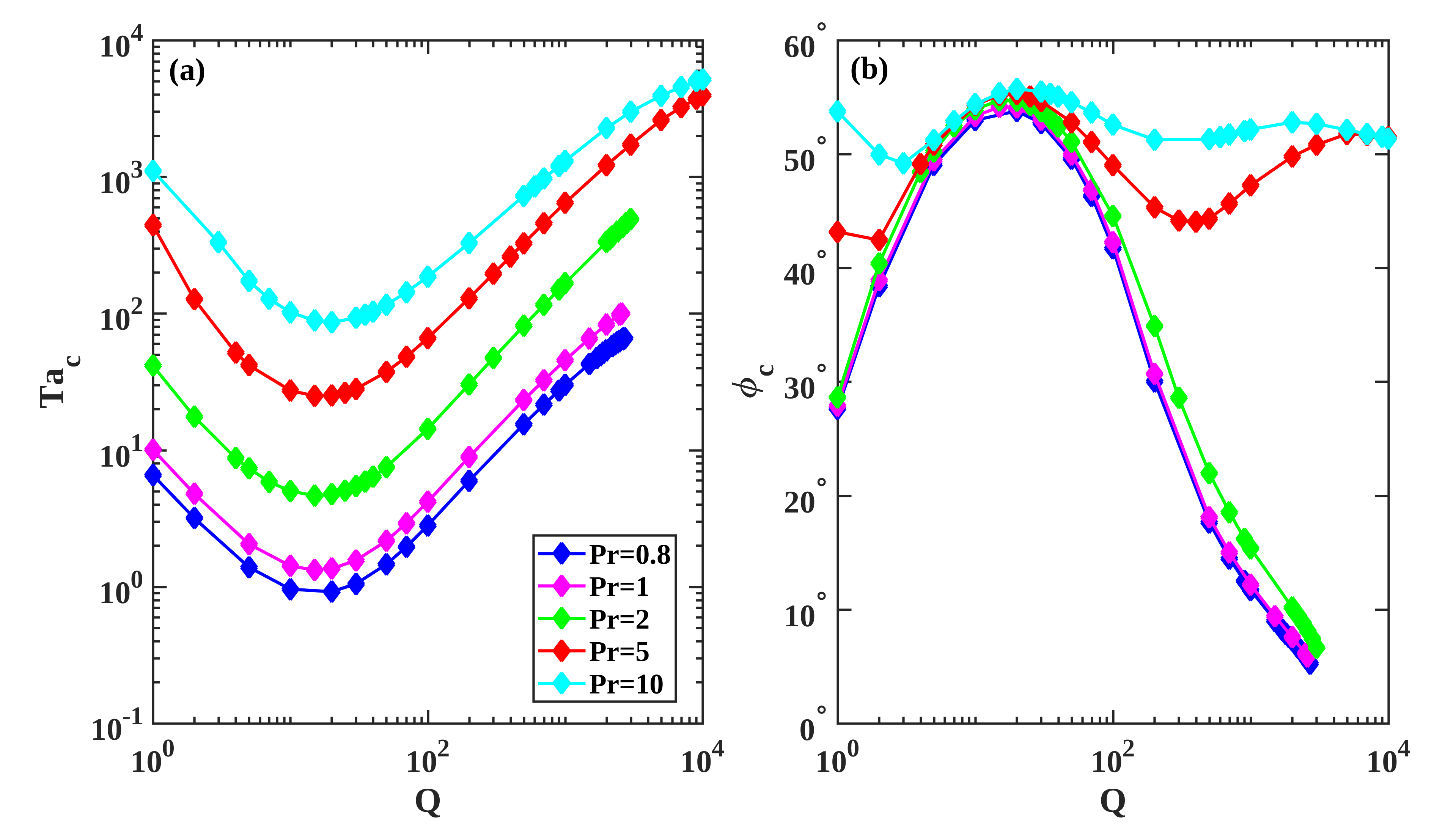}
\caption{(a) Critical Taylor number ($\mathrm{Ta}_c$) and (b) critical angle ($\phi_c$) of the onset of KLI as a function of $\mathrm{Q}$ for smaller Prandtl numbers.}
\label{smallPr_Q}
\end{figure}
However, unlike the high Prandtl number fluids, for lower Prandtl number fluids, the graph of $p_2$ show the divergence near $\phi = 0^\circ$ as previously pointed out in~\cite{ponty:POF_1997,Clune:PRE1993}. FIG.~\ref{smallPr_Q50} shows effect of $\mathrm{Ta}$ on the graphs of the growth rate as a function of the angle $\phi$. It is observed that the graph of $p_2$ after showing divergence near $\phi = 0^\circ$, it becomes negative for lower value of $\phi$ and subsequently moves up to become positive again for higher $\phi$. The divergence of $p_2$ near $\phi \rightarrow 0$, is associated with the small angle instability (SAI) reported earlier~\cite{ponty:POF_1997,cox:2000}. On the other hand, the positive part of the graph of $p_2$ for higher value of $\phi$ is associated with the K\"{u}ppers-Lortz instability which is the topic of interest in this paper.  The effect of $\mathrm{Pr}$ on the SAI will be investigated in detail in a separate paper. 

\begin{figure}[h!]
\includegraphics[height=!,width=0.5\textwidth]{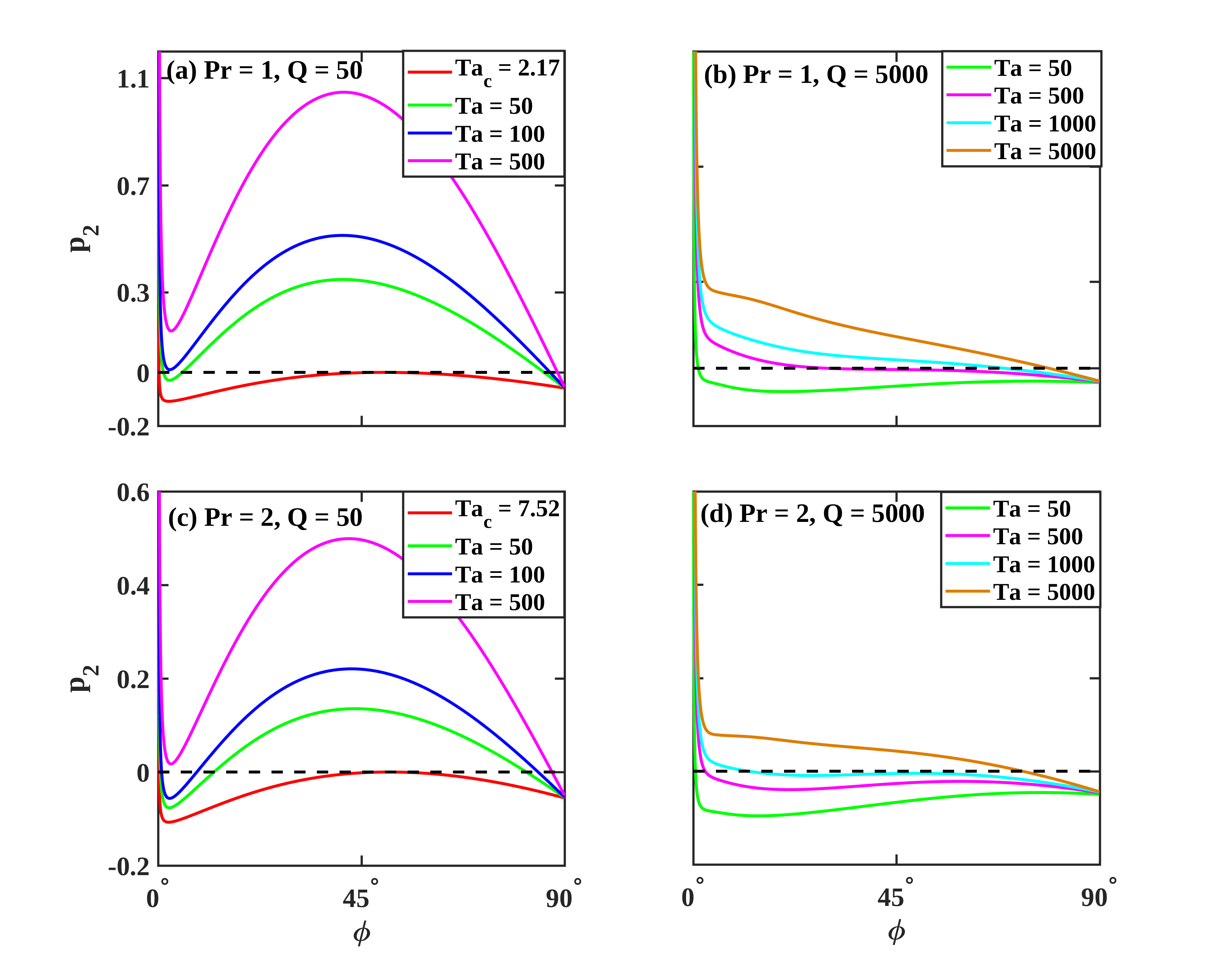}
\caption{The effect of $\mathrm{Q}$ on the variation of $p_2$ for $\mathrm{Pr} = 1$ (first row) and $2$ (second row) with four different $\mathrm{Ta}$ in each case. The values of $\mathrm{Q}$ taken in (a), (b), (c) and (d) are $50$, $5000$, $50$ and $5000$ respectively. In (b) and (d), curves remain below the horizontal line once it reaches there indicating the complete suppression of KLI.}
\label{smallPr_Q}
\end{figure}

As $\mathrm{Pr}$ is reduced further,  not only the drastic decrease in $\mathrm{Ta}_c$ is observed in the absence of magnetic field, but a substantial lowering of the local minima for relatively weaker magnetic field is observed.  For larger magnetic field, $\mathrm{Ta}_c$ consistently increases with $\mathrm{Q}$ like high Prandtl number fluids. Interestingly,  for $\mathrm{Pr} \leq 2$,  the critical angle for the onset of KLI,  first increases with $\mathrm{Q}$ to reach a local maxima and then continuously decreases to a small value before it ceased to exist for $\mathrm{Q} > 30.$ The cessation of the KLI for higher $\mathrm{Q}$ can be understood from the effect of $\mathrm{Ta}$ on the graph of $p_2$ as a function of the angle $\phi$ shown in the FIG.~\ref{smallPr_Q}. From the figure it is observed that both for $\mathrm{Pr} = 1$ and $2$ as the strength of the magnetic field is increased, the graph of $p_2$ once comes below the zero line, it remains in the negative part for higher $\phi$ for all considered values of $\mathrm{Ta}$. Therefore, whenever the strength of the magnetic field is increased beyond a critical value, SAI takes over and KLI ceases to exist. 
This is an interesting effect of the external horizontal magnetic field  on the KLI of low Prandtl number fluids.

\section{Conclusions}\label{sec5:conclusion}
We have theoretically investigated the onset of K\"uppers-Lortz instability in rotating Rayleigh-B\'enard convection with free-slip boundary conditions in the presence of horizontal external magnetic field using weakly nonlinear theory. A wide region of the parameter space ($0.8 \leq \mathrm{Pr} < \infty$, $0 < \mathrm{Ta} \leq 10^4$ and $0 \leq \mathrm{Q} \leq 10^4$) has been considered for the investigation in the vanishingly small magnetic Prandtl number limit. Through the theoretical investigation, different ranges of $\mathrm{Pr}$ have been identified where KLI and SAI are manifested. In this paper, we focus on the KLI and investigate the effect of external magnetic field on it.   

In the infinite $\mathrm{Pr}$ limit, in the absence of external horizontal magnetic field, the weakly nonlinear analysis presented in this paper closely reproduces the original KLI results~\cite{kuppers:1969}. In this limit, the external magnetic field inhibits the KLI and significantly enhances the stability region of the 2D-rolls flow patterns in the parameter space. Thus, as the strength of the external magnetic field is increased in this case, the critical Taylor number ($\mathrm{Ta}_c$) for the onset of KLI consistently enhanced and the corresponding critical angle $\phi_c$ decreased. For  larger $\mathrm{Pr} (\leq 500)$, similar effect of external magnetic field,  on the onset of KLI is observed.  

For the fluids with Prandtl numbers in the range $50 \leq \mathrm{Pr} < 500$, an interesting change in the effect of magnetic field on the onset of KLI is observed. As the Prandtl number gradually decreased in this range in the absence of the magnetic field, the critical Taylor number ($\mathrm{Ta}_c$) for the onset of KLI also decreases together with the corresponding critical angle $\phi_c$. However, unlike the large to infinite Prandtl number fluids, small magnetic field is found to promote KLI. In this case, $\mathrm{Ta}_c$ is found to decrease with $\mathrm{Q}$ together with the corresponding $\phi_c$. With the increase of $\mathrm{Q}$ starting from a small value, $\mathrm{Ta}_c$ and corresponding $\phi_c$ eventually attain local minima. On further increase of $\mathrm{Q}$,  $\mathrm{Ta}_c$ continuously increases i.e. KLI is largely inhibited by the external magnetic field. The corresponding $\phi_c$ also decreases for large external magnetic field after reaching a local maxima near $\mathrm{Q}\sim 10^2$.  

As the Prandtl number is decreased further ($\mathrm{Pr} < 50$), in the absence of the external magnetic field, not only the critical Taylor number for the onset of KLI decreases substantially, but also the divergence of the growth rate $p_2$ as $\phi \rightarrow 0$ is observed which indicates the appearance of small angle instability in the system and we plan to investigate it in detail in a future work. The K\"{u}ppers-Lortz  instability along with the small  angle instability are found to coexist in this range of Prandtl numbers. In this range of $\mathrm{Pr}$ also, weak magnetic field is found to promote KLI along with SAI. Stronger magnetic field, interestingly, inhibit KLI and promote SAI. For very low Prandtl number fluids ($\mathrm{Pr} \leq 2$), the KLI is completely suppressed for higher $\mathrm{Q}$. The interesting nontrivial effects of the external horizontal magnetic field on the KLI revealed from the present study may lead to the investigation of KLI in the finite magnetic Prandtl number fluids. 


\begin{center}
{\bf ACKNOWLEDGEMENTS}
\end{center}
SM, SS and PP acknowledge the supports from the CSIR, India [Award No. 09/973(0024)/2019-EMR-I], UGC, India [Award No. 191620126754] and SERB, India [Grant No. MTR/2017/000945] respectively. The authors thank L. Sharma, S. Hansda, P. P. Gopmandal and S. De  for insightful comments on the manuscript and helping in preparing some of the figures.

\begin{center}
{\bf APPENDIX}
\end{center}
The coefficients used in the equations (\ref{Y1_u0}) - (\ref{p2_ex}) and (\ref{finitePr_w1}) - (\ref{p2_finitePr}) are defined as follows: 
\begin{widetext}
\begin{eqnarray}
A_\pm &=& -\frac{\mathrm{R_c}k_c^2(1 \pm \mathrm{cos}\phi)}{\Delta_\pm}\left[\frac{\pi (1\mp\mathrm{cos}\phi)}{\pi ^2+k_c^2} \pm 2P \right],\\
P &=& \frac{\pi \sqrt{\mathrm{Ta}}~ \mathrm{sin}\phi}{4(\pi^2+k_c^2)^2} - \frac{\pi \sqrt{\mathrm{Ta}}~ \mathrm{sin}\phi}{4((\pi^2+k_c^2)^2+\mathrm{Q}k_2^2)},\\
B_\pm &=& \frac{2\pi \sqrt{\mathrm{Ta}}(2k_c^2(1 \pm \mathrm{cos}\phi)+4\pi^2)}{(2k_c^2(1 \pm \mathrm{cos}\phi)+4\pi^2)^2+\mathrm{Q}k_2^2}A_\pm,\\
C_\pm &=& \frac{-\pi(1\mp \mathrm{cos}\phi)\{2k_c^2(1\pm\mathrm{cos}\phi)+4\pi^2\}\{(2k_c^2(1\pm\mathrm{cos}\phi)+4\pi^2)^2+\mathrm{Q}k_2^2\}}{2(\pi^2+k_c^2)\{2k_c^2(1\pm\mathrm{cos}\phi)+4\pi^2\}\Delta_\pm} \\ \nonumber &&- \frac{4\pi^3\mathrm{Ta}(1\mp\mathrm{cos}\phi)\{2k_c^2(1\pm \mathrm{cos}\phi)+4\pi^2\}^2}{(\{2k_c^2(1 \pm \mathrm{cos}\phi)+4\pi^2\}^2+\mathrm{Q}k_2^2)[2(\pi^2+k_c^2)(2k_c^2(1 \pm \mathrm{cos}\phi)+4\pi^2)]\Delta_\pm} \mp \frac{1+\frac{2\mathrm{R_c}k_c^2(1\pm \mathrm{cos}\phi)P}{\Delta_\pm}}{2k_c^2(1\pm \mathrm{cos}\phi)+4\pi^2}\\
\Delta_\pm &=& \{2k_c^2(1\pm \mathrm{cos}\phi)+4\pi^2\}\{(2k_c^2(1\pm \mathrm{cos}\phi)+4\pi^2)^2+\mathrm{Q}k_2^2\}-2k_c^2\mathrm{R_c}(1\pm \mathrm{cos}\phi) \\ \nonumber && +
 \frac{4\pi^2\mathrm{Ta}(2k_c^2(1\pm \mathrm{cos}\phi)+4\pi^2)}{(2k_c^2(1 \pm \mathrm{cos}\phi)+4\pi^2)^2+\mathrm{Q}k_2^2},\\
a_\pm &=& \frac{-\pi (\pi^2+k_c^2)\sqrt{\mathrm{Ta}}~\mathrm{sin}\phi\{\mathrm{cos}\phi \pm (\pi^2+k_c^2)\} -\pi^3 \mathrm{Ta}~\mathrm{cos2}\phi +\pi^3\mathrm{Ta}(1+\mathrm{sin^2}\phi)-\pi^3(\pi^2+k_c^2)\mathrm{\sqrt{Ta}}~\mathrm{sin}\phi(\mathrm{cos}\phi\mp \mathrm{sin}\phi)}{4\{(\pi^2+k_c^2)^2+\mathrm{Q}k_2^2\}}  \nonumber \\ &+& \frac{1}{4}\left[\pi(\pi^2+k_c^2) + \pi\sqrt{\mathrm{Ta}}~\mathrm{sin}\phi(\mathrm{\cos}\phi\pm 1) +\frac{\pi^3 \sqrt{\mathrm{Ta}}~\mathrm{sin}\phi(\mathrm{cos}\phi\mp 1)}{\pi^2 + k_c^2}\right],\\
b_\pm &=& \frac{2\pi^3 \sqrt{\mathrm{Ta}}~\mathrm{sin}\phi\mathrm{cos}\phi+\pi k_c^2\sqrt{\mathrm{Ta}}~\mathrm{sin}\phi(\mathrm{cos}\phi\pm 1)}{4(\pi^2 +k_c^2)}+\frac{\pi(\pi^2 +k_c^2)\mathrm{sin^2}\phi}{2} \nonumber \\ &+& \left[\frac{\pi^2\mathrm{Ta}~\mathrm{sin^2}\phi(1 \pm \mathrm{cos}\phi)-\pi \sqrt{\mathrm{Ta}}(\pi^2 + k_c^2)\mathrm{sin}\phi\{2\pi^2 \mathrm{cos}\phi+k_c^2(\mathrm{cos}\phi \pm 1\}+\pi^3 \mathrm{Ta}~\mathrm{sin^2}\phi(1+\mathrm{cos}\phi \mp \mathrm{2cos}\phi)}{4\{(\pi^2 +k_c^2)^2+\mathrm{Q}k_2^2\}}\right],\\
c_\pm &=& \mathrm{Pr}^{-1}(a_\pm +b_\pm)(2k_c^2\pm 2k_c^2\mathrm{cos}\phi+4\pi^2), ~d_\pm = \mathrm{R_c}\left[\{\frac{\pi(1\mp \mathrm{cos}\phi)}{2(\pi^2+k_c^2)}\pm P\}2k_c^2(1\pm \mathrm{cos}\phi)\right], ~e_\pm = \frac{1}{\Delta_\pm}(-c_\pm -d_\pm),\\
N_1 &=& \frac{\pi^2 \sqrt{\mathrm{Ta}}(\pi^2 +k_c^2)}{2\{(\pi^2+k_c^2)^2+\mathrm{Q}k_2^2\}} + \frac{\pi^2 \sqrt{\mathrm{Ta}}}{2(\pi^2 +k_c^2)},
~g_\pm = -N_1\mathrm{Pr^{-1}}4k^4\frac{(1\pm \mathrm{cos}\phi)^3}{\{2k_c^2(1\pm \mathrm{cos}\phi)\}^2+\mathrm{Q}k_2^2},\\
h_\pm &=& \frac{2\pi \sqrt{\mathrm{Ta}}\{2k_c^2(1\pm \mathrm{cos}\phi)+4\pi^2\}}{\{2k_c^2(1\pm \mathrm{cos}\phi)+4\pi^2 \}^2+\mathrm{Q}k_2^2}e_\pm,~f_\pm = -\left[\frac{\frac{\pi (1\mp \mathrm{cos}\phi)}{2(\pi^2+k_c^2)}\pm P -e_\pm}{2k_c^2(1\pm \mathrm{cos}\phi)+4\pi ^2}\right],\\
q_\pm &=& \frac{k_2g_\pm}{2k_c^2(1\pm \mathrm{cos}\phi)},~r_\pm = \frac{k_2 h_\pm}{2k_c^2(1\pm \mathrm{cos}\phi)}, ~\mathrm{and}~s_\pm = \frac{2\pi (k_1\pm k)e\pm}{2k_c^2(1\pm \mathrm{cos}\phi)}.
\end{eqnarray}
\end{widetext}

%

%
%
\end{document}